\documentclass[10pt,twocolumn]{IEEEtran}
\usepackage{cite}
\usepackage{amsmath,amssymb,amsfonts}
\usepackage{graphicx}
\usepackage{textcomp}
\usepackage{xcolor}
\def\BibTeX{{\rm B\kern-.05em{\sc i\kern-.025em b}\kern-.08em
    T\kern-.1667em\lower.7ex\hbox{E}\kern-.125emX}}
\usepackage{pgfplots}
\pgfplotsset{compat=newest}
\usetikzlibrary{external}
\usepackage{svg}
    \usepackage{ragged2e}
    \usetikzlibrary{shapes.geometric, arrows.meta, positioning, calc, fit, backgrounds}
\usepackage{floatrow}
\usepackage{tabu}
\usepackage{multirow}
\usepackage{amsthm}
\usepackage[utf8]{inputenc}
\usepackage[english]{babel}
\theoremstyle{remark}
\usetikzlibrary{calc}
\usepackage[export]{adjustbox}
\theoremstyle{plain} 

\usepackage{multicol}
\usepackage{epsfig}
\usetikzlibrary{arrows.meta}
\usepackage{epstopdf}
\usepackage{float}
\usepackage{perpage}
\MakeSorted{figure}
\MakeSorted{table}
\usepackage[normalem]{ulem}
\usepackage{array}
\usepackage{graphicx}
\usepackage{amsfonts}
\usepackage{amssymb}
\usepackage{soul}

\usepackage{enumitem}							
\usepackage[english]{babel}
\usepackage[utf8]{inputenc}
\usepackage[linesnumbered,ruled,vlined]{algorithm2e}
\usepackage{upgreek}
\usepackage{bm} 
\usepackage{hyperref}
\usepackage{float}
\floatstyle{plaintop}
\restylefloat{table}
\hypersetup{linktocpage} 
\hypersetup{
colorlinks,
citecolor=black,
filecolor=black,
linkcolor=black,
urlcolor=black
}
\usepgfplotslibrary{fillbetween} 
\usepackage{tikz}
\usetikzlibrary{arrows.meta}
\usepackage{pifont}
\usepackage{makecell}
\usepackage{algorithmic}

\usepackage{pgfplots}
\usetikzlibrary{shapes.geometric, arrows.meta, positioning, calc, fit, backgrounds}
\tikzset{
    decision/.style = {
        diamond, 
        draw, 
        fill=gray!20, 
        text width=2.5cm, 
        text centered, 
        inner sep=0pt, 
        aspect=2
    },
    base/.style={rectangle, rounded corners, draw=black!70, text centered, minimum height=0.7cm, fill=white}
}

\usepackage{cite}
\DeclareMathAlphabet\mathbfcal{OMS}{cmsy}{b}{n}
\ifCLASSINFOpdf
\else
\fi
\usepackage{pgfplots}
\pgfplotsset{compat=newest}
\usepackage{tikz}

\usepackage{array}
\usepackage{fixltx2e}

\usepackage[font=footnotesize]{caption} 
\usepackage{subcaption}
\usepackage{fix-cm}
\usepackage{comment}
\usepackage{pgfplotstable}
\usepackage{pgfplots}
\usepgfplotslibrary{colormaps}
\usepgfplotslibrary{patchplots}

\usepackage{tikz}
\usepackage{pgfplots}
\pgfplotsset{compat=newest}
\usepgfplotslibrary{external}

\usepackage{mathtools}
\usepackage{pifont}

\usepackage{mathtools}
\usepackage{pifont}

\usepackage{dsfont}

\DeclareRobustCommand{\blacksolidline}{\tikz[baseline=-0.5ex]\draw[black, thick] (0,0) -- (0.5,0);}
\DeclareRobustCommand{\blackdashed}{\tikz[baseline=-0.5ex]\draw[black, thick, dashed] (0,0) -- (0.5,0);}
\DeclareRobustCommand{\bluedashed}{\tikz[baseline=-0.5ex]\draw[blue!60, thick, dashed] (0,0) -- (0.5,0);}
\DeclareRobustCommand{\bluesolidline}{\tikz[baseline=-0.5ex]\draw[blue!60, thick] (0,0) -- (0.5,0);}

\begin{document}
\bstctlcite{IEEEexample:BSTcontrol}

\title{
Radio Imaging and Resource Allocation in Frugal Multistatic D-MIMO ISAC Systems
}
\author{Sauradeep Dey, Musa~Furkan~Keskin,~\textit{Senior Member,~IEEE}, Dario~Tagliaferri,~\textit{Member,~IEEE}, Gonzalo~Seco-Granados,~\textit{Fellow,~IEEE}, Henk~Wymeersch, \textit{Fellow,~IEEE}
\thanks{{This work was partly supported by the SNS JU project 6G-DISAC under the EU’s Horizon Europe research and innovation program under Grant Agreement No 101139130, the Swedish Research Council (VR) through the project 6G-PERCEF under Grant 2024-04390, the Spanish R+D project PID2023-152820OB-I00, and the AGAUR-ICREA Academia Program.}}
	\thanks{S. Dey, M. F. Keskin and H. Wymeersch are with the Department of Electrical Engineering, Chalmers University of Technology, Gothenburg, Sweden (emails: \{deysa, furkan, henkw\}@chalmers.se)
}	
\thanks{D. Tagliaferri is with the Department of Electronics, Information and Bioengineering, Politecnico di Milano, 20133, Milano, Italy (e-mail: dario.tagliaferri@polimi.it).}
\thanks{G. Seco-Granados is with the Department of Telecommunications and Systems Engineering, Universitat Autonoma de Barcelona, Spain (email:  gonzalo.seco@uab.cat).}\vspace{-35pt}}

\maketitle

\begin{abstract}
Emerging integrated sensing and communication (ISAC) systems based on distributed MIMO (D-MIMO) enable radio imaging by exploiting spatial diversity across multiple access points (APs). However, joint operation of sensing and communication in such systems introduces fundamental challenges due to mutual interference between the communication and sensing  signals. In this paper, we propose a downlink D-MIMO ISAC framework that allocates orthogonal subcarriers to sensing and communication to eliminate the inter-function interference. We consider a phase-coherent architecture in which single-antenna APs simultaneously serve communication user equipments (UEs) while  constructing a reflectivity image of the environment. 
We develop a two-timescale resource allocation framework that minimizes the entropy of the reconstructed image subject to communication spectral efficiency (SE) constraint. {The proposed design follows a communication-centric resource allocation policy, where imaging is supported using the resources left after satisfying the communication requirement.} 
The long-timescale optimization (LTO) determines AP modes (transmission or reception) and subcarrier assignment, including the partitioning of the subcarriers between communication and sensing functions and the allocation of sensing subcarriers among transmit APs, based on synthetic scenarios with random UE and target distributions. The short-timescale optimization (STO) dynamically adapts the communication-sensing power-splitting factor to preserve the SE constraint based on the current scenario.
Numerical evaluations demonstrate that the proposed orthogonal subcarrier allocation achieves a superior sensing–communication trade-off compared with conventional superposition scheme, where the sensing and communication signals share the same subcarriers. Finally, we evaluate coherent and non-coherent imaging receivers and identify the synchronization regimes under which each approach  achieves superior imaging and localization performance.
\end{abstract}

\vspace{-5pt}
\section{Introduction}
\vspace{-3pt}
Distributed multiple-input multiple-output (D-MIMO) has emerged as a promising architecture for sixth-generation (6G) wireless systems~\cite{Demir2021Foundations}. Multiple geographically distributed access points (APs) operating in a phase-coherent manner effectively form a large virtual antenna array~\cite{TENTU_ISAC,Sofie_1}. This improves the   communication through coherent transmission and macro-diversity gains~\cite{Bjornson2020Scalable}, while enabling high-resolution sensing by coherent combination of observations from spatially distributed viewpoints~\cite{TENTU_ISAC,Tagliaferri_cooperative,Sofie_1}. These capabilities make phase-coherent D-MIMO an attractive platform for integrated sensing and communication (ISAC)~\cite{Nuria_Proc_IEEE_2024,Buzzi_CFISAC,Sofie_1,TENTU_ISAC}.

Early D-MIMO ISAC research mainly focused on target localization, where the objective is to estimate the target position, velocity, or orientation. Recently, attention has shifted toward radio imaging~\cite{FanLiu_imaging,IIAC_3D_imaging,Alkhateeb23_imaging_comm,Tagliaferri_cooperative}, which reconstructs a two- or three-dimensional reflectivity map of the environment. Such environmental awareness is essential for emerging applications such as digital twins, autonomous systems, and extended reality. 
In parallel, existing D-MIMO ISAC works have investigated resource-allocation strategies for jointly optimizing sensing and communication performance. These works formulate resource-allocation problems to maximize communication performance subject to sensing constraints~\cite{AP_MODE_EE,AP_MODE_ELFIATOURE}, sensing performance subject to communication constraints~\cite{Alkhateeb_cellfreeISAC,AP_MODE_SIFAN,TWO_TIME_SCALE}, or optimize a weighted combination of sensing and communication metrics~\cite{Li_BF_Design_grouping_CFISAc,AP_MODE_SUN}. Typical optimization variables include AP modes (transmit/receive),  beamforming, power allocation, and waveform design~\cite{AP_MODE_EE,AP_MODE_ELFIATOURE,Alkhateeb_cellfreeISAC,AP_MODE_SIFAN,TWO_TIME_SCALE,Li_BF_Design_grouping_CFISAc,AP_MODE_SUN,Lou_BF_design_CFISAC,Masouros2025_distributedISAC,AP_MODE_YAN}.

{The sensing performance in the aforementioned works is primarily evaluated using sensing signal-to-noise ratio (SNR)~\cite{Alkhateeb_cellfreeISAC,AP_MODE_EE,AP_MODE_SIFAN,AP_MODE_SUN,TWO_TIME_SCALE}, probability of detection~\cite{Lou_BF_design_CFISAC}, or target position  Cramér-Rao lower bound (CRLB)~\cite{AP_MODE_YAN,Sofie_1}  based metrics. Although sensing SNR is a useful indicator of signal quality, it does not necessarily translate into improved target detection or localization performance. Similarly, works using the probability of detection are often limited to simple single-target scenarios~\cite{Lou_BF_design_CFISAC}. Moreover, CRLB-based formulations require prior knowledge of the target locations~\cite{AP_MODE_YAN}. Consequently, these metrics are insufficient for characterizing sensing performance in multi-target environments with unknown target locations. 
A separate modeling limitation in several of these works is that communication and sensing are described using different channel models.
In particular, communication links are typically modeled using stochastic fading channels, whereas sensing links rely on LoS-only channels~\cite{AP_MODE_ELFIATOURE,Masouros2025_distributedISAC,Alkhateeb_cellfreeISAC,AP_MODE_EE,AP_MODE_SIFAN,AP_MODE_SUN}. This simplifies the analysis but overlooks the physical coupling between sensing and communication through the same propagation environment. Finally, the works discussed above~\cite{AP_MODE_EE,AP_MODE_ELFIATOURE,Alkhateeb_cellfreeISAC,AP_MODE_SIFAN,TWO_TIME_SCALE,Li_BF_Design_grouping_CFISAc,AP_MODE_SUN,Lou_BF_design_CFISAC,Masouros2025_distributedISAC,AP_MODE_YAN} focus on target localization or tracking, where only the target parameters are estimated. They do not reconstruct the surrounding environment and therefore cannot provide a spatial map of the~scene. }

{In contrast to target localization, applications such as digital twins, autonomous systems, and extended reality require a richer understanding of the environment. This has motivated growing interest in radio imaging for D-MIMO ISAC systems, where the goal is to reconstruct a reflectivity map of the coverage area rather than estimate only a few target parameters~\cite{Liu_survey,moreira2013tutorial}. The distributed geometry of D-MIMO is well suited for this task. Spatially separated APs observe the scene from different viewpoints, which improves spatial resolution and robustness to occlusions~\cite{Gonzalez-Prelcic2025,manzoni2024wavefield,Tagliaferri_cooperative}. Recent works have therefore investigated imaging in distributed wireless systems~\cite{IIAC_3D_imaging,Tong2022_occlusion,li2024networked,zhi2025nearfield_distributed,yang2025nearfield,yang2025illumination,gao2025covariance}.
However, many of these approaches rely on multi-antenna APs or large antenna arrays to obtain spatial resolution, which increases the hardware complexity.

From an algorithmic perspective, the aforementioned works employ compressed sensing, message passing, or covariance-based imaging methods. Compressed sensing and message passing require iterative optimization and can become computationally expensive for large imaging grids. Covariance-based methods require multiple temporal snapshots to estimate a reliable covariance matrix, which is less suitable for dynamic environments. To overcome these limitations, back-projection (BP) imaging has attracted significant attention~\cite{tagliaferri2025integratingphasecoherentmultistaticimaging}. BP provides a linear, single-step imaging framework that reconstructs the environment with low computational complexity and only a single snapshot~\cite{tagliaferri2025integratingphasecoherentmultistaticimaging}. 
Among BP-based approaches, our previous work~\cite{tagliaferri2025integratingphasecoherentmultistaticimaging} investigated downlink multistatic imaging in phase-coherent D-MIMO networks. This framework superposed AP-specific sensing waveforms onto the communication signals, allowing both functions to share the entire system bandwidth. However, sharing the same subcarriers introduced mutual interference between sensing and communication, degrading both the imaging quality and the communication performance. In addition, the framework assumed partial prior knowledge of the target locations by restricting them to a predefined region of interest. The sensing signals were beamformed toward this region to reduce interference with the communication~signal. }

{In this paper, we address the above limitations by considering a frugal downlink D-MIMO ISAC  architecture with single-antenna APs, and orthogonal subcarrier allocation. The design allocates disjoint subcarriers to sensing and communication, thereby eliminating  sensing-communication interference. We develop coherent and non-coherent BP imaging receivers to reconstruct the reflectivity map of the coverage area. The AP transmit/receive modes, subcarrier allocation, and sensing-communication power split are optimized to minimize the reconstructed-image entropy while satisfying communication constraints. The subcarrier allocation includes both the partitioning between communication and sensing functions and the allocation of sensing subcarriers among transmit APs. Unlike CRLB-, detection probability-, or sensing SNR-based metrics, entropy directly measures image quality in multi-target environments without requiring prior target locations~\cite{ENT_1,Ent_2,ENT_3}.}

The main contributions of this paper are listed as follows:
\begin{itemize}
    \item \textbf{Orthogonal subcarrier-based D-MIMO ISAC framework and entropy-based sensing evaluation}: We propose a frugal multistatic downlink D-MIMO ISAC framework with single-antenna APs and orthogonal frequency-domain multiplexing of sensing and communication signals. {The sensing subcarriers are further allocated disjointly across the transmit APs, making the AP-specific echoes separable at the receive APs.} 
    Sensing is performed via image reconstruction of the coverage area, which is  used to detect and localize multiple point targets. To evaluate the sensing performance, we employ the image entropy. It quantifies the overall spatial concentration of the reconstructed image without any prior knowledge about the targets.
    
   \item \textbf{Two-timescale optimization algorithm}: We develop an  optimization framework to minimize the expected image entropy while ensuring communication sum-spectral-efficiency (SE) constraints, by selecting AP transmit/receive modes, sensing subcarriers, and sensing-communication power split. 
   {This reflects a communication-centric design in which imaging is performed using the resources available after meeting the communication requirement.}
   To ensure computational feasibility and practical implementation, the optimization is decoupled across two timescales. A long-timescale optimization (LTO) operates offline to determine the discrete AP transmit/receive modes and sensing subcarrier assignments based on one or more synthetic scenarios, with random target and UE distributions. Subsequently, a short-timescale optimization (STO) operates online to dynamically adapt the continuous sensing-communication power splitting factor according to instantaneous channel conditions.
    
    \item \textbf{Sensing-communication trade-offs and robustness analysis}: We numerically demonstrate that the proposed orthogonal subcarrier allocation achieves a superior sensing-communication trade-off compared to the superposed coding scheme in~\cite{tagliaferri2025integratingphasecoherentmultistaticimaging}. The proposed framework consistently attains lower image entropy over a wide range of communication sum-SE budgets, and improved multi-target localization accuracy, measured by the generalized optimal sub-pattern assignment (GOSPA) metric~\cite{GOSPA}. We also show that maximizing the frequency diversity of each bistatic transmitter-receiver pair through interleaved subcarrier allocation provides better sensing performance than clustered allocation, where the sensing subcarriers occupy contiguous frequency bands. Finally, we evaluate the robustness of the proposed framework to imperfect AP synchronization and investigate coherent and non-coherent imaging performance.
\end{itemize}
\vspace{-8pt}
\section{System Model and Problem Statement}\vspace{-0pt}
\begin{figure}[t]
    \centering
    \begin{subfigure}[h]{\linewidth}
        \centering
        \input{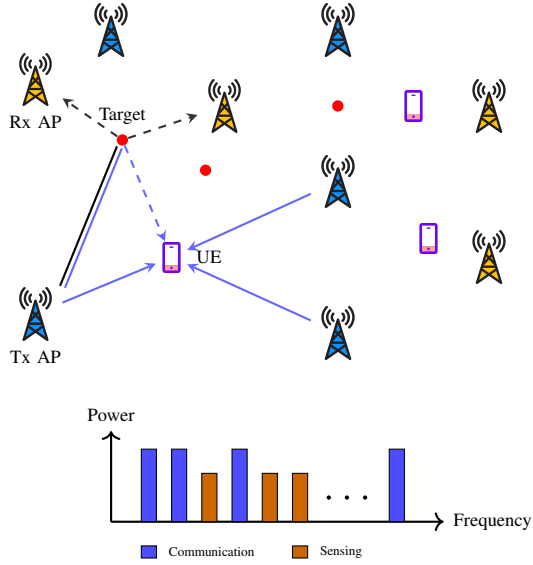}
        \end{subfigure}
        \vspace{-7pt}
           \caption{System model of a D-MIMO ISAC network with orthogonal subcarrier allocation. Arbitrarily distributed single-antenna APs operate in transmit and receive modes, serving communication UEs and sensing point targets distributed over the coverage area. The illustrated links highlight the dual functionality: direct communication links to UEs (\bluesolidline); sensing illumination paths from transmit APs to targets (\blacksolidline), reflected paths from targets to receive APs (\blackdashed); and the reflected paths from targets to UEs (\bluedashed). The lower plot shows the frequency-domain resource allocation, where sensing and communication signals occupy orthogonal subcarriers.\vspace{-2pt}}
    \label{fig:system_model}
\end{figure}

{In this section we present the downlink D-MIMO ISAC system model, along with the corresponding transmit and receive signal models, and the channel model.}

\vspace{-7pt}
\subsection{Scenario}\vspace{-2pt}
{We consider a D-MIMO system, as illustrated in Fig. \ref{fig:system_model}, where $G$ single-antenna APs are deployed arbitrarily over a coverage area $\mathcal{A}$. The system simultaneously serves $L$ single-antenna UEs in downlink and senses multiple targets. Target sensing is performed by first generating a high-resolution image of the coverage area, and then detecting the peaks to estimate the target locations. 
We assume that the APs are phase-synchronized and the $g$-th AP is located at $\mathbf{p}_g = [x_g, y_g]^\top$, while the $\ell$-th UE and $q$-th target are located at $\mathbf{u}_\ell = [u_{x,\ell}, u_{y,\ell}]^\top$ and $\mathbf{r}_{q}=[r_{x,q}, r_{y,q}]^\top$, respectively. 
We split the total set of $G$ APs between $N$ transmit APs, indexed by the set $\mathcal{T}$, and $M = G - N$ receive APs, indexed~by~set~$\mathcal{R}$.}

{The APs employ orthogonal frequency division multiplexing (OFDM) waveform to transmit downlink data to the UEs and pilot signals for target sensing, over $S$ subcarriers with subcarrier spacing $\Delta_f$. 
The $S$ subcarriers are partitioned into $S_{\rm sen}$ sensing subcarriers, and $S_{\rm com} = S - S_{\rm sen}$ communication subcarriers, enforcing orthogonality between communication and sensing. The sensing and communication subcarriers are indexed by the sets $\mathcal{S}_{\rm sen}$ and $\mathcal{S}_{\rm com}$, respectively. The sensing subcarriers are further split among the $N$ transmit APs into  disjoint sets $\{\mathcal{S}_{n,\rm sen}\}_{n\in \mathcal{T}}$, where $\mathcal{S}_{n,\rm sen}$, with cardinality $\kappa_n$, denotes the set of subcarriers allocated to the $n$-th AP for sensing. {The limiting case where $\kappa_n=1$, $\forall n\in \mathcal{T}$, i.e. each transmit AP uses one subcarrier for sensing, is referred to as \textit{phase only sensing (POSE)}}. The $S_{\rm com}$ communication subcarriers are jointly used by all transmit APs for data~transmission.}

\vspace{-5pt}
\subsection{Signal and Channel Model}\label{subsec:signalmodel}\vspace{-2pt}
{The transmit signal vector over the $s$-th subcarrier is denoted by $\mathbf{x}[s]\in\mathbb{C}^{N\times 1}$ and is given as follows:}
\begin{align}
\begin{split}
    \mathbf{x}[s] = \begin{dcases}
       \mathbf{F}_{\rm com}[s]\;\mathbf{a}_{\rm com}[s] & \text{for} \; s \in \mathcal{S}_{\rm com} \\
       \mathbf{a}_{\rm sen}[s] & \text{for} \; s \in \mathcal{S}_{\rm sen}
    \end{dcases}
\end{split}.
\end{align}
{The vector $\mathbf{a}_{\rm com}[s]\in\mathbb{C}^{L\times 1}$ contains the data symbols of the $L$ UEs transmitted on the $s$-th subcarrier. These symbols are precoded by $\mathbf{F}_{\rm com}[s]\in\mathbb{C}^{N\times L}$, which is normalized as $\|\mathbf{F}_{\rm com}[s]\|_F^2=L$. The data symbols are assumed to be uncorrelated and equally powered, i.e.,$\mathbb{E}\{\mathbf{a}_{\rm com}[s]\mathbf{a}_{\rm com}^{\rm H}[s]\}
= \frac{\alpha P}{L S_{\rm com}}\mathbf{I}_L$,
where $P$ is the total transmit power and $\alpha\in[0,1]$ is the fraction of power allocated to communication.
The vector $\mathbf{a}_{\rm sen}[s] \in \mathbb{C}^{N\times 1}$ represents the sensing pilot signal transmitted over the $s$-th subcarrier, with $\mathbb{E}\{\|\mathbf{a}_{\rm sen}[s]\|^2\}= (1-\alpha)P/S_{\rm sen}$. To ensure orthogonal allocation of sensing subcarriers across the transmitting APs, we set $\mathbf{a}_{\rm sen}[s] =\sqrt{(1-\alpha)P/S_{\rm sen}}  \mathbf{\mathds{1}}_{n^{(s)}}$ where $\mathbf{\mathds{1}}_{n^{(s)}} \in \mathbb{B}^{N \times 1}$ is a one-hot vector with a single non-zero entry at index $n^{(s)}\in\mathcal{T}$.
This ensures that only the $n^{(s)}$-th transmitter transmits the $s$-th sensing subcarrier. No precoding is applied to the sensing pilots, as the transmitting APs use orthogonal sensing~subcarriers.}

{The communication channel between $N$ transmitting APs and $L$ UEs on the $s$-th subcarrier $(s \in \mathcal{S}_{\rm com})$ is denoted by $\mathbf{G}[s] \in\mathbb{C}^{L\times N}$, and can be decomposed as follows~\cite{RS_LOC}}
\begin{equation}\label{eq:comm_channel_mat}
    \mathbf{G}[s] = \mathbf{G}_{\rm los}[s] +  \mathbf{G}_{\rm scat}[s] \,,
\end{equation}
{where $\mathbf{G}_{\rm los}[s]\in\mathbb{C}^{L\times N}$ captures the line of sight (LoS) component of the channel, and $\mathbf{G}_{\rm scat}[s]\in \mathbb{C}^{L\times N}$ models the scattering of the transmit signal by the targets in the coverage area. The $(\ell, n)$-th entry of each matrix corresponds to the respective channel component between the $n$-th transmit AP and $\ell$-th UE, and are given as follows~\cite{tagliaferri2025integratingphasecoherentmultistaticimaging,tornielli2025enabling_nlos_imaging}}
\begin{align}
        &\!\!\!\!\!\![\mathbf{G}_{\rm los}[s]]_{\ell n} = \frac{\lambda_0 e^{-j\phi_\ell} e^{-j\frac{2\pi}{c} (f_0 + s \Delta_f) d(\mathbf{p}_n,\mathbf{u}_\ell)}}{4\pi  d(\mathbf{p}_n,\mathbf{u}_\ell)} \label{eq:comm_channel_mat_LOS}  \\
         &\!\!\!\!\!\!\![\mathbf{G}_{\rm scat}[s]]_{\ell n}\!\!=\!\! \int\!\! \frac{\xi[\mathbf{r}] \lambda_0 e^{\!-j\phi_\ell} e^{\!-j\frac{2\pi}{c} (f_0 + s \Delta_f) \left(d(\mathbf{p}_n,\mathbf{r}) + d(\mathbf{r},\mathbf{u}_\ell)\right) }}{(4 \pi)^{\frac{3}{2}}d(\mathbf{p}_n,\mathbf{r})  d(\mathbf{r},\mathbf{u}_\ell) \, } \text{d}\mathbf{r}\!\!\!\label{eq:comm_channel_mat_scat}
\end{align}
{where $\lambda_0$ and $f_0$ are the carrier wavelength and frequency respectively, and $c$ is the speed of light. The $\ell$-th UE has an unknown phase offset $\phi_\ell$ with respect to the AP network~\cite{POLO}.
The term $\xi[\mathbf{r}]d\mathbf{r}$ represents the complex reflectivity of location $\mathbf{r}$ in the coverage area $\mathcal{A}$. We use  $d(\mathbf{x},\mathbf{y})=\|\mathbf{x}-\mathbf{y}\|_2$ to denote the distance between the coordinates $\mathbf{x}$ and $\mathbf{y}$.}

{The sensing channel from the $N$ transmit APs to the $M$ receive APs on subcarrier $s\in\mathcal{S}_{\rm sen}$ is denoted by $\mathbf{H}[s]\in\mathbb{C}^{M\times N}$, with its $(m,n)$-th element given by~\cite{tornielli2025enabling_nlos_imaging}} 
\begin{align}\label{eq:csensing_channel_mat}
    [\mathbf{H}[s]]_{mn} = \int \frac{\xi[\mathbf{r}] \lambda_0 e^{-j\frac{2\pi}{c} (f_0 + s \Delta_f) \left(d(\mathbf{p}_n,\mathbf{r}) + d(\mathbf{r},\mathbf{p}_m)\right) }}{(4 \pi)^{\frac{3}{2}}d(\mathbf{p}_n,\mathbf{r})  d(\mathbf{r},\mathbf{p}_m)} \text{d} \mathbf{r}.
\end{align}
{Since the AP locations are fixed and known a-priori, the LoS paths between the transmitting and receiving APs can be accurately determined and their contribution, similar to the works in~\cite{AP_MODE_ELFIATOURE,Alkhateeb_cellfreeISAC,AP_MODE_SIFAN,TWO_TIME_SCALE}, is assumed to be mitigated. }

{The  received signal at the UEs on the $s$-th subcarrier $(s\in\mathcal{S}_{\rm com})$ is given as}
\begin{align}\label{eq:Rx_comm_signal}
    \mathbf{y}[s] = \mathbf{G}[s] \, \mathbf{F}_{\rm com}[s] \, \mathbf{a}_{\rm com}[s] + \mathbf{n}[s]
\end{align}
{where $\mathbf{n}[s] \sim \mathcal{CN}(\mathbf{0},\sigma_n^2 \mathbf{I}_L)$ is the additive Gaussian noise (AWGN), assumed to be uncorrelated among UEs and subcarriers. 
The receive signal at the $M$ receive APs over the $s$-th subcarrier $(s\in\mathcal{S}_{\rm sen})$ is given as }
\begin{equation}\label{eq:Rx_sensing_signal}
    \mathbf{z}[s] = \mathbf{H}[s] \, \mathbf{a}_{\rm sen}[s] + \mathbf{w}[s]
\end{equation}
{where $\mathbf{w}[s] \sim \mathcal{CN}(\mathbf{0},\sigma_n^2 \mathbf{I}_M)$ is the AWGN, assumed to be uncorrelated among the receiving APs and subcarriers.}

\vspace{-5pt}
\subsection{Problem Statement}\vspace{-2pt}
The objective of the considered D-MIMO ISAC system is to simultaneously support communication for the UEs and sensing of targets in the environment. Given the received sensing signals $\{\mathbf{z}[s]\}_{s\in\mathcal{S}_{\rm sen}}$ in \eqref{eq:Rx_sensing_signal}, target sensing is performed by reconstructing an image of the environment, followed by peak detection to estimate the target locations. To quantify the image quality, we use entropy \cite{ENT_1,Ent_2,ENT_3}, detailed in Section \ref{subsec:entropy}. Unlike CRLB- or detection-probability based metrics, it does not require prior knowledge of the target locations and can easily be applied to complex multi-target scenarios.

Since sensing and communication functions share the available transmit power and bandwidth, and are constrained through AP transmit/receive mode selection, they are inherently coupled. The system design therefore requires joint optimization of the AP transmit/receive modes, the sensing/communication subcarrier allocation $\{\mathcal{S}_{\rm sen}/\mathcal{S}_{\rm com}\}$, and the power allocation factor $\alpha$ to maximize the sensing performance, quantified by the image entropy, while satisfying the required communication quality, quantified by a sum-SE constraint. To achieve this, we develop a two-timescale optimization framework, where the AP modes and sensing subcarriers are optimized offline in long-timescale based on synthetic scenarios, and the power-factor is updated dynamically in the short-timescale using the instantaneous channel~conditions.

\vspace{-7pt}
\section{Receiver-Side Processing: Imaging and Target Localization}\vspace{-2pt}
{This section presents the receiver-side processing framework for sensing. We first describe the imaging principle and consider both coherent and non-coherent image reconstruction approaches. We then discuss target detection and localization from the reconstructed image.}
\vspace{-7pt}
\subsection{Imaging of the Environment}\vspace{-2pt}
We reconstruct the image of the coverage area using BP over a uniformly spaced grid of pixels $\{\mathbf{r}\}$ covering the entire sensing region, with an inter-pixel spacing of $0.025$~m~\cite{BP_1}. BP is a widely used imaging benchmark and provides an estimate of the reflectivity $\xi[\mathbf{r}]$ at each pixel. Although nonlinear imaging techniques can improve image quality, they incur significantly higher computational complexity~\cite{NON_LINEAR_IMAGING}.

We first construct a pairwise image for each transmit-receive AP pair. {The one-hot structure of the sensing transmit signal $\mathbf{a}_{\rm sen}[s]$ ensures that, for $s\in\mathcal{S}_{n,\rm sen}$, the received sensing signal $[\mathbf{z}[s]]_m$ at the $m$-th AP contains only the echo associated with the AP pair $(n,m)$. For each candidate pixel $\mathbf{r}$, this echo is matched to the expected  propagation phase along the path $\mathbf{p}_n\rightarrow\mathbf{r}\rightarrow\mathbf{p}_m$ and then coherently combined across the sensing subcarriers assigned to the $n$-th transmit AP $\mathcal{S}_{n,\rm sen}$. The resulting pairwise image is given by}
\begin{align}\label{eq:image_nc_1}
I_{mn}[\mathbf{r}]=\sum_{s\in\mathcal{S}_{n,\rm sen}}[\mathbf{z}[s]]_m\psi_{mns}[\mathbf{r}],
\end{align}
where $\psi_{mns}[\mathbf{r}]=e^{j\big(\frac{2\pi}{c}(f_0+s\Delta_f)(d(\mathbf{p}_n,\mathbf{r})+d(\mathbf{r},\mathbf{p}_m))
\big)}$ is the matched filter kernel.
The final image is obtained by combining the images from different transmit-receive AP pairs. Depending on whether the frequency and phase coherence across AP pairs is exploited, two image reconstruction methods are possible: coherent imaging and non-coherent imaging.

\subsubsection{Coherent Imaging}\label{sec:coherent_imaging}
{Coherent imaging exploits phase coherence across all APs. The pairwise images are therefore added coherently to form the final image~\cite{tagliaferri2025integratingphasecoherentmultistaticimaging}}
\begin{align}\label{eq:image}
I[\mathbf{r}]=\sum_{m\in\mathcal{R}}\sum_{n\in\mathcal{T}}
I_{mn}[\mathbf{r}].
\end{align}
{At the target location, the phase-compensated contributions from different AP pairs add constructively, producing a strong and narrow peak. {At non-target locations, the matched contributions are associated with negligible reflectivity, resulting in low-amplitude image background after coherent combination.} Consequently, coherent imaging improves image contrast and enhances target detection and localization accuracy. This coherent gain is particularly important in the POSE regime, where each transmit AP uses only a single sensing subcarrier. Although each AP pair provides no carrier frequency diversity, different transmit APs operate on different sensing subcarriers. Coherent combination of the pairwise images exploits the spatial and frequency diversity for reliable image~reconstruction.}

\subsubsection{Non-coherent Imaging}\label{sec:noncoherent_imaging} 
{Non-coherent imaging does not exploit phase synchronization across different AP pairs. The image is obtained by combining the magnitudes of the single images in \eqref{eq:image_nc_1}, as
\begin{align}\label{eq:image_nc}
|I_{\rm nc}[\mathbf{r}]|^2=\sum_{m\in\mathcal{R}}\sum_{n\in\mathcal{T}}|I_{mn}[\mathbf{r}]|^2.
\end{align}
{This formulation avoids the need for network-wide phase synchronization, but sacrifices the gain obtained by coherent combination across different AP pairs. In the POSE regime ($\kappa=1$), each transmit AP occupies only a single sensing subcarrier, so each AP pair contributes only one complex observation, eliminating the intra-pair frequency diversity. As a result, non-coherent imaging becomes ineffective.}

\vspace{-5pt}
\subsection{Target Localization} \vspace{-2pt}
{The target localization process converts the reconstructed image into a set of discrete target location estimates $\{\widehat{\mathbf{r}}_q\}_{q=1}^{\widetilde{Q}}$\footnote{{In the target-localization analysis, we consider point targets without explicit clutter. If clutter is present, clutter suppression can be applied to the image before target detection; the design of such processing is outside the scope of this paper.}}, where $\widetilde{Q}$ denotes the number of detected targets. Target detection is first performed using a constant false alarm rate (CFAR) detector~\cite{CFAR_JALIL}.  
The target peaks returned by CFAR provide coarse estimates on the discretized imaging grid. The resulting set of detected target locations is defined as $\overline{\mathcal{Q}}=\big\{\overline{\mathbf{r}}_q\big\}_{q=1}^{\widetilde{Q}}.$}

To improve the localization accuracy beyond the grid resolution, each detected peak is further refined using gradient ascent on the image intensity function.
Starting from the coarse estimate $\overline{\mathbf{r}}_q$, the target location is refined by applying gradient ascent to the continuously evaluated BP image intensity $I[\mathbf{r}]$~as 
\begin{align}
\widehat{\mathbf{r}}_q=\mathrm{GA}\big(|I[\mathbf{r}]|^2,
\overline{\mathbf{r}}_q\big),
\end{align}
{where $\mathrm{GA}(f(\mathbf{x}),\overline{\mathbf{x}})$ denotes the gradient ascent procedure applied to $f(\mathbf{x})$ using $\overline{\mathbf{x}}$ as the initialization point~\cite{BOYD_CV_OPT}. The set of final target estimates is defined as $\widehat{\mathcal{Q}}=\{\widehat{\mathbf{r}}_q\}_{q=1}^{\widetilde{Q}}$.}

\section{Transmitter-Side Processing: \\ Two-Timescale Resource Allocation}\label{sec_opt_res_aloc}
{This section focuses on the transmitter-side design of the considered D-MIMO ISAC system. The objective is to determine: \textit{(i)} the AP transmit/receive modes, \textit{(ii)} the sensing/communication subcarrier allocation $\{\mathcal{S}_{\rm sen}/\mathcal{S}_{\rm com}\}$, and \textit{(iii)} the power allocation factor $\alpha$, to maximize the sensing performance while satisfying the communication requirement. We first introduce the sensing and communication performance metrics, and then formulate the resource allocation problem. }

\vspace{-5pt}
\subsection{Performance Metrics}\vspace{0pt}
{We use the image entropy and sum-SE to evaluate the sensing and communication performance, respectively.}

\subsubsection{Entropy}\label{subsec:entropy}
The entropy of the image quantifies the spatial concentration of the reconstructed image energy. Lower entropy indicates that the image energy is concentrated around sharp and well-localized target responses, whereas higher entropy indicates increased energy dispersion caused by sidelobes and reconstruction artifacts~\cite{ENTROPY, tagliaferri2025integratingphasecoherentmultistaticimaging}. In contrast to sensing metrics, such as CRLB, detection probability and sensing SNR \cite{Alkhateeb_cellfreeISAC,AP_MODE_EE,AP_MODE_SIFAN,AP_MODE_SUN,TWO_TIME_SCALE,Lou_BF_design_CFISAC, AP_MODE_YAN}, image entropy quantifies the image quality in a generic multi-target scenario, without any prior target knowledge\footnote{A low-entropy image does not necessarily imply accurate target reconstruction, as image energy may remain concentrated around incorrect target locations. Therefore, target detection and localization performance must also be evaluated using metrics such as GOSPA~\cite{GOSPA}, as done in Section~\ref{sec:Num_results}.}\cite{ENT_1,Ent_2,ENT_3}. 
To evaluate the entropy, we first normalize the image amplitudes, so that it can be interpreted as a probability distribution over the coverage area: $\overline{I}[\mathbf{r}]=|I[\mathbf{r}]|^2/\sum\limits_{\mathbf{r}'\in \mathcal{A}}|I[\mathbf{r}']|^2$, where $\mathcal{A}$ represents the coverage area. The entropy is then calculated as follows:
\begin{align}
    \mathcal{E}[I] = -\sum\limits_{\mathbf{r}\in\mathcal{A}} \overline{I}[\mathbf{r}]\log_2(\overline{I}[\mathbf{r}]).
\end{align}

\subsubsection{Sum-SE}\label{subsec:sumrate}
The communication performance is quantified using the sum-SE, obtained by summing the SEs of the $L$ UEs over the communication subcarriers and normalizing by the total system bandwidth, as follows:
\begin{align}
    &\mathrm{R} =\frac{1}{S} 
    \sum\limits_{\ell=1}^L\sum\limits_{s\in\mathcal{S}_{\rm com}}\log_2(1+\mathrm{SINR}_{\ell,s}), \text{ where }\notag\\[-3pt]
    &\!\!\!\!\mathrm{SINR}_{\ell,s} = \frac{\frac{\alpha P}{LS_{\rm com}} \Big|\sum\limits_{n\in\mathcal{T}} [\mathbf{G}[s]]_{\ell n}[\mathbf{F}_{\rm com}[s]]_{n\ell}\Big|^2}{\sum\limits_{\ell'\neq \ell}\frac{\alpha P}{LS_{\rm com}} \Big|\sum\limits_{n\in\mathcal{T}}[\mathbf{G}[s]]_{\ell n}[\mathbf{F}_{\rm com}[s]]_{n\ell'}\Big|^2+{\sigma_n^2}}.\!\!\!\label{eq:SINR_def}
\end{align}

\vspace{-7pt}
\subsection{Problem Formulation}\label{sec:problem}
 {We begin by introducing the following optimization variables: \textit{(i)} the AP mode vector $\mathbf{b} \in \mathbb{B}^{G \times 1}$, where $[\mathbf{b}]_n = 1$ indicates that the $n$-th AP is a transmitter, and $[\mathbf{b}]_n = 0$ that it is a receiver; \textit{(ii)} the subcarrier allocation matrix $\mathbf{S} \in \mathbb{B}^{G \times S}$, where $[\mathbf{S}]_{gs} = 1$ indicates that the $g$-th AP uses the $s$-th subcarrier for sensing; and \textit{(iii)} the power splitting factor $\alpha \in [0,1]$. The joint resource allocation problem can be formulated as a image entropy minimization problem subject to a communication sum-SE constraint, as follows:}
\begin{subequations}\label{eqn:P1}
\begin{alignat}{2}
&&\textbf{P1: }\underset{\mathbf{b},\,\mathbf{S},\,\alpha}{\mbox{ min }}\;&
  \mathcal{E}(\mathbf{b}, \mathbf{S}, \alpha) 
  \label{eqn:P1_obj}\\
&&\text{ s.t. }\;
  &[\mathbf{b}]_g \in\{0, 1\},\; \forall\, g,
  \label{eqn:P1_b}\\
  &&&[\mathbf{S}]_{gs}\in\{0, 1\},\; \forall\, g,\; 
  \forall\, s, 
  \label{eqn:P1_S}\\
   &&& \alpha\in[0,1],
  \label{eqn:P1_alpha}\\
  &&&[\mathbf{S}]_{gs} \leq [\mathbf{b}]_g,\; \forall\, g,\; 
  \forall\, s,
  \label{eqn:P1_coupling}\\
  &&&\sum\limits_{g=1}^{G} [\mathbf{S}]_{gs}\leq 1,\; 
  \forall\, s,
  \label{eqn:P1_ortho}\\
  &&& \mathrm{R}(\mathbf{b}, \mathbf{S}, \alpha) \geq\gamma.
  \label{eqn:P1_rate}
\end{alignat}
\end{subequations}
The constraints \eqref{eqn:P1_b} and \eqref{eqn:P1_S} enforce the variables  $\mathbf{b}$ and $\mathbf{S}$ to take Boolean values, and the constraint \eqref{eqn:P1_alpha} restricts the power-splitting factor $\alpha$ between $[0,1]$. The constraint \eqref{eqn:P1_coupling} ensures that sensing subcarriers are assigned only to transmitting APs, and the constraint \eqref{eqn:P1_ortho} ensures that each transmit AP uses orthogonal sensing subcarriers. The constraint \eqref{eqn:P1_rate} ensures the sum-SE meets the minimum threshold $\gamma$. The threshold $\gamma$ is set as a fraction of the maximum achievable sum-SE $\mathrm{R}_{\max}$, i.e., $\gamma = \beta \mathrm{R}_{\max}$. Here $\mathrm{R}_{\max}$ is obtained by setting all $G$ APs as transmitters, $\mathbf{S}=\mathbf{0}$, and $\alpha=1$.  The parameter $\beta\in(0,1)$ specifies the normalized sum-SE budget, i.e., the fraction of $\mathrm{R}_{\rm max}$ that must be guaranteed. Consequently, problem \textbf{P1} seeks the best sensing configuration while satisfying the communication requirement. {The formulation applies to both coherent and non-coherent imaging by evaluating $\mathcal{E}(\mathbf{b},\mathbf{S},\alpha)$ using the corresponding reconstructed image in \eqref{eq:image} or \eqref{eq:image_nc}, respectively.}

The joint optimization of $\{\mathbf{b},\mathbf{S},\alpha\}$ in \eqref{eqn:P1} results in a mixed-integer non-convex problem. The complexity arises from the binary variables $\{\mathbf{b},\mathbf{S}\}$ and the non-convex entropy objective. {The objective is non-convex because the entropy is computed from the normalized reconstructed-image intensity, where the image intensity itself is a function of AP modes, sensing subcarrier allocation, and power factor through the received AP signals $\mathbf{z}[s]$.}  Consequently, standard convex optimization techniques cannot be directly applied. Finding the global optimum, therefore,  requires  exhaustive search over the combinatorial solution space, whose size grows exponentially with the network size. 
This makes solving \eqref{eqn:P1} at every transmission snapshot computationally impractical for large-scale D-MIMO networks. Furthermore, updating the AP transmit/receive modes or reallocating sensing subcarriers requires network-wide coordination, resulting in significant backhaul overhead~\cite{CF_OVERHEAD}. In contrast, the power allocation factor $\alpha$ is a continuous scalar that can be rapidly updated according to the instantaneous channel conditions~\cite{TWO_TIME_SCALE}.

{To address these challenges, we adopt a two-timescale optimization framework consisting of an LTO and an STO stage.  The LTO stage is performed offline using synthetic scenarios with randomly distributed UEs and targets. For each normalized communication budget $\beta$, it jointly optimizes $\mathbf{b}$, $\mathbf{S}$, and $\alpha$ to minimize the expected image entropy, while satisfying the sum-SE constraint. The optimized pair $(\mathbf{b}^*,\mathbf{S}^*)$ is stored as a codebook entry for online use. During the STO stage, $(\mathbf{b}^*,\mathbf{S}^*)$ initializes the online optimization, and $\alpha$ is recomputed according to the instantaneous channel realization to satisfy the sum-SE constraint. If the sum-SE constraint is not satisfied even with $\alpha=1$, the  $(\mathbf{b}^*,\mathbf{S}^*)$ are further refined to restore feasibility. Since STO starts from a near-optimal LTO solution, the online updates incur only limited computational overhead. The implementation details of the LTO and STO stages are presented next.}

\vspace{-10pt}
\subsection{Long-Timescale Optimization (LTO)}\label{sec:LTO}
{The LTO stage is performed offline using a collection of synthetic scenarios, generated by randomly distributing UEs and targets in the coverage area. For each scenario, the corresponding {sensing and communication} channels are generated based on the system model. Using these channels, we formulate the expected image entropy and  sum-SE metrics for each synthetic scenario. The variables $\mathbf{b}$, $\mathbf{S}$, and $\alpha$ are then jointly optimized to minimize the maximum expected image entropy across all scenarios, while satisfying a constraint on the average sum-SE across the synthetic scenarios for a given value of $\beta$.}

\subsubsection{Scenario Generation for LTO}\label{sec:scenario_generation}
{The LTO stage is performed using a set of $K$ synthetic scenarios. In the $k$-th scenario, $Q$ point targets and $L$ UEs are randomly dropped in the coverage area. Let $\{\widetilde{\mathbf{u}}_{\ell}^{(k)}\}_{\ell=1}^{L}$ and $\{\widetilde{\mathbf{r}}_{q}^{(k)}\}_{q=1}^{Q}$ denote the UE and target locations, respectively. Using these positions, the synthetic communication and sensing channel matrices,  $\widetilde{\mathbf{G}}^{(k)}[s]$ and $\widetilde{\mathbf{H}}^{(k)}[s]$ $\forall s\in\mathcal{S}$, are generated according to \eqref{eq:comm_channel_mat} and \eqref{eq:csensing_channel_mat}, respectively. The resulting channel realizations for all $K$ scenarios are pre-computed and stored for use~during~the~LTO stage.}
\subsubsection{ Sensing and Communication Metrics for LTO}\label{sec:expected_image}
{To evaluate the sensing performance offline, we simulate the received sensing observations using the synthetic channels $\widetilde{\mathbf{H}}^{(k)}[s]$. Specifically, for a given transmit-receive AP pair $(n,m)$ on subcarrier $s \in \mathcal{S}_{n,\rm sen}$, the synthetic observation using \eqref{eq:Rx_sensing_signal} for the $k$-th scenario is modeled as:}
\begin{equation}
[\widetilde{\mathbf{z}}^{(k)}[s]]_m = \sqrt{\frac{(1-\alpha)P}{S_{\rm sen}}} [\widetilde{\mathbf{H}}^{(k)}[s]]_{mn}
+ [\mathbf{w}^{(k)}[s]]_m,
\end{equation}
{where $[\mathbf{w}^{(k)}[s]]_m \sim \mathcal{CN}(0, \sigma_n^2)$ is the AWGN at the $m$-th receive AP.
The resulting image intensity $\widetilde{I}^{(k)}[\mathbf{r}; \mathbf{b},\mathbf{S},\alpha]$ is computed by substituting  $[\widetilde{\mathbf{z}}^{(k)}[s]]_m$ in \eqref{eq:image}. 
The image intensity can be decomposed into a signal-dependent component and a noise-dependent component as follows}
\begin{equation}\label{eqn:image_decomp}
\widetilde{I}^{(k)}[\mathbf{r}; \mathbf{b},\mathbf{S}, \alpha] = \widetilde{I}_{\rm sig}^{(k)}[\mathbf{r}; \mathbf{b},\mathbf{S}, \alpha] + W^{(k)}[\mathbf{r}],
\end{equation}
{where the signal component $\widetilde{I}_{\rm sig}^{(k)}[\mathbf{r}; \mathbf{b},\mathbf{S}, \alpha]$, and the noise component $W^{(k)}[\mathbf{r}]$ are given as}
\begin{align}
  &\!\!\!\!\!\widetilde{I}_{\rm sig}^{(k)}[\mathbf{r}; \mathbf{b},\mathbf{S}, \alpha]\! =\! \sqrt{\frac{(1\!-\!\alpha)P}{S_{\rm sen}}}
  \!\!\!\!\sum_{m\in\mathcal{R}(\mathbf{b})}
  \sum_{\substack{n\in\mathcal{T}(\mathbf{b})\\s\in\mathcal{S}_{n,\rm sen}}}
\!\!\!
  [\widetilde{\mathbf{H}}^{(k)}[s]]_{mn}
  \psi_{mns}[\mathbf{r}]\nonumber\\[-5pt]
  &\hspace{55pt}\triangleq \sqrt{\frac{(1\!-\!\alpha)P}{S_{\rm sen}}} \widetilde{J}^{(k)}(\mathbf{r};\mathbf{b},\mathbf{S})\!\!\!\label{eqn:I_sig}\\
  \label{eqn:W_noise}
  &W^{(k)}[\mathbf{r}] = \sum\limits_{m \in \mathcal{R}(\mathbf{b})} \sum\limits_{n \in \mathcal{T}(\mathbf{b})} 
  \sum\limits_{s \in \mathcal{S}_{n,\rm sen}} 
  [\mathbf{w}^{(k)}[s]]_m\, \psi_{mns}[\mathbf{r}].
\end{align}
{Since $\mathbf{w}^{(k)}[s]$ is independent across subcarriers and receive APs, and the sensing subcarrier sets $\{\mathcal{S}_{n,\rm sen}\}_{n \in \mathcal{T}}$ are disjoint, the noise component $W^{(k)}[\mathbf{r}]$ has zero mean, i.e., $\mathbb{E}\{W^{(k)}[\mathbf{r}]\}=0$ and its variance is given by}
\begin{equation}\label{eqn:noise_power}
  \mathbb{E}\{|W^{(k)}[\mathbf{r}]|^2\} = \sigma_n^2 M S_{\rm sen} \,.
\end{equation}
{The expected image intensity for the $k$-th scenario is given as:}
\begin{equation}
  \mathbb{E}\{|\widetilde{I}^{(k)}[\mathbf{r}; \mathbf{b},\mathbf{S}, \alpha]|^2\} = 
  {\frac{(1\!-\!\alpha)P}{S_{\rm sen}}} \big|\widetilde{J}^{(k)}(\mathbf{r};\mathbf{b},\mathbf{S})\big|^2 + \sigma_n^2 M(\mathbf{b}) S_{\rm sen}.\nonumber
\end{equation}
{The corresponding expected entropy for the $k$-th scenario is}
\begin{align}\label{eq:exp_ent}
\!\!\!\!    \widetilde{\mathcal{E}}^{(k)}(\mathbf{b},\mathbf{S}, \alpha)
    \!=\! -\!\sum\limits_{\mathbf{r}\in\mathcal{A}}\! \breve{I}^{(k)}[\mathbf{r}; \mathbf{b},\mathbf{S},\alpha]
      \log_2\Big(\!\breve{I}^{(k)}[\mathbf{r}; \mathbf{b},\mathbf{S}, \alpha]\!\Big)\!.\!\!\!
\end{align}
{where $\breve{I}^{(k)}[\mathbf{r}; \mathbf{b},\mathbf{S},\alpha]= { \mathbb{E}\{|\widetilde{I}^{(k)}[\mathbf{r}; \mathbf{b},\mathbf{S},\alpha]|^2\}}/\sum\limits_{\mathbf{r'}\in\mathcal{A}}\mathbb{E}\{|\widetilde{I}^{(k)}[\mathbf{r}';$ $ \mathbf{b},\mathbf{S},\alpha]|^2\}$. This serves as the objective for the LTO.} To enforce the communication requirement in  LTO, we evaluate the system sum-SE for each synthetic scenario, denoted by $\widetilde{\mathrm{R}}^{(k)}(\mathbf{b}, \mathbf{S},\alpha)$. It is obtained by substituting the instantaneous channels $\mathbf{G}[s]$ by the synthetic channels~$\widetilde{\mathbf{G}}^{(k)}[s]$~in~\eqref{eq:SINR_def}. 
\subsubsection{Problem Formulation for LTO}
{We now formulate the LTO problem, which jointly optimizes $\{\mathbf{b}, \mathbf{S},\alpha\}$ to minimize the maximum expected image entropy across all synthetic scenarios, while satisfying the constraint on the {average} sum-SE across the scenarios. Optimizing the power factor $\alpha$ during the LTO stage ensures that the AP mode and subcarrier allocation are obtained under the best possible power allocation. However, only the optimized pair $(\mathbf{b}^*,\mathbf{S}^*)$ is stored as the codebook entry for a given   budget $\beta$. During the STO stage, $\alpha$ is re-optimized according to the instantaneous channel realizations, while using $(\mathbf{b}^*,\mathbf{S}^*)$ to initialize the optimization.}

{The LTO problem is cast as follows}
\begin{subequations}\label{eqn:P2_long_timescale}
\begin{alignat}{2}
&&\textbf{P2: }\underset{\mathbf{b},\,\mathbf{S},\alpha}{\mbox{ min }}\;&\underset{k\in\mathcal{K}}{\mbox{max }}
  \widetilde{\mathcal{E}}^{(k)}(\mathbf{b}, \mathbf{S},\alpha) 
  \label{eqn:P2_obj}\\
&&\text{ s.t. }\;
  &[\mathbf{b}]_g \in\{0, 1\},\; \forall\, g,
  \label{eqn:P2_b}\\
  &&& \alpha\in(0,1)\label{eqn:P2_alpha}\\
  &&&[\mathbf{S}]_{gs}\in\{0, 1\},\; \forall\, g,\; 
  \forall\, s, 
  \label{eqn:P2_S}\\
  &&&[\mathbf{S}]_{gs} \leq [\mathbf{b}]_g,\; \forall\, g,\; 
  \forall\, s,
  \label{eqn:P2_coupling}\\
  &&&\sum\limits_{g=1}^{G} [\mathbf{S}]_{gs}\leq 1,\; 
  \forall\, s,
  \label{eqn:P2_ortho}\\
  &&& \widetilde{\mathrm{R}}_{\rm avg}(\mathbf{b}, \mathbf{S},\alpha) \geq\widetilde{\gamma}_{\rm avg}.
  \label{eqn:P2_rate}
\end{alignat}
\end{subequations}
{Here $\widetilde{\mathrm{R}}_{\rm avg}(\mathbf{b}, \mathbf{S},\alpha)=1/K\sum_{k=1}^K\widetilde{\mathrm{R}}^{(k)}(\mathbf{b}, \mathbf{S},\alpha)$ represents the average sum-SE across $K$ scenarios. The average sum-SE threshold $\widetilde{\gamma}_{\rm avg}$ is set as $\beta /K\sum_{k=1}^K\widetilde{\mathrm{R}}^{(k)}_{\max}$, where $\widetilde{\mathrm{R}}^{(k)}_{\max} = \widetilde{\mathrm{R}}^{(k)}(\mathbf{b}=\mathbf{1}_G, \mathbf{S}=\mathbf{0},\alpha=1)$ is the maximum achievable sum-SE for the $k$-th synthetic scenario.} To solve \textbf{P2}, we adopt a greedy-search-based optimization technique. We first study the AP mode selection, sensing subcarrier allocation, and power factor optimization  individually. Using insights from them, we develop a joint LTO strategy for solving \textbf{P2}. 

\begin{algorithm}[t]
\DontPrintSemicolon
\SetAlgoLined
\KwIn{AP positions $\{\mathbf{p}_g\}$, synthetic channels $(\widetilde{\mathbf{G}}^{(k)}[s],\widetilde{\mathbf{H}}^{(k)}[s])$  $\forall s\,\,\forall k$, sum-SE budget~$\beta$\;}
\KwOut{$\mathbf{b}^*$, $\kappa^*$, $\mathbf{S}^*, \alpha^*$}
\BlankLine
{Set {$\widetilde{\gamma}_{\rm avg}=\beta/K\sum\limits_{k=1}^{K}\widetilde{\mathrm{R}}^{(k)}_{\max}$}\;
Initialize $\mathbf{b}^{(0)}$ with $N^{(0)}=G-1$ and $M^{(0)}=1$ \;
Set $F^{(0)}\gets\infty$\;
}
\While{$\mathrm{true}$}
{
\For{$g = 1, \ldots, G$}{
        Generate $\mathbf{b}^{(l,g)}$ by flipping $g$-th entry of $\mathbf{b}^{(l)}$\;
        \If{ 
        {$\widetilde{\mathrm{R}}_{\rm avg}(\mathbf{b}^{(l,g)},\kappa^{(l,g)}\!=\!1,\alpha\!=\!1)\!\geq\!\widetilde{\gamma}_{\rm avg}$}}
        {{Compute $\kappa_{\rm max}^{(l,g)}$ using~\eqref{eqn:kappa_max} with $\mathbf{b}^{(l,g)}$\;
        \For{$\kappa\in\{1,\ldots,\kappa_{\rm max}^{(l,g)}\}$}{
            Set $\mathbf{S}_\kappa^{(l,g)}$ using \eqref{eqn:interleaved} with $\kappa$\;
            Calculate $\alpha_\kappa^{(l,g)}$ using \eqref{eqn:opt_alpha_LTO}\;
            Set $F_\kappa^{(l,g)}\!\!=\!\underset{k\in\mathcal{K}}{\max}\,\widetilde{\mathcal{E}}^{({k})}(\mathbf{b}^{(l,g)}\!\!,\mathbf{S}_\kappa^{(l,g)}\!,\alpha_\kappa^{(l,g)})$\;
            }
        $ F^{(l,g)}\gets \underset{\kappa}{\min}\,F_\kappa^{(l,g)}$
        $\kappa^{(l,g)}\!\gets\underset{\kappa}{\arg\min}\,F_\kappa^{(l,g)}$, $\mathbf{S}^{(l,g)}\!\gets\!\mathbf{S}_{\kappa^{(l,g)}}^{(l,g)}$, $\alpha^{(l,g)}\!\gets\!\alpha_{\kappa^{(l,g)}}^{(l,g)}$\;}
            
        \Else{Set $F^{(l,g)} \gets +\infty$\;} }}
    Select $g^* = \underset{g}{\arg\min}\, F^{(l,g)}$\;
    \If{$F^{(l)}-F^{(l,g^*)}>\epsilon$}{Update $\mathbf{b}^{(l+1)} \gets \mathbf{b}^{(l,g^*)}$,
    $\kappa^{(l+1)} \gets \kappa^{(l,g^*)}$,
    $\mathbf{S}^{(l+1)} \gets \mathbf{S}^{(l,g^*)}$, {$\alpha^{(l+1)} \gets \alpha^{(l,g^*)}$,}
    $F^{(l+1)}\gets F^{(l,g^*)}$, $l \gets l+1$\;}
    \Else{\textbf{break}}}
Set $(\mathbf{b}^*, \kappa^*, \mathbf{S}^*, {\alpha^*}) \gets (\mathbf{b}^{(l)}, \kappa^{(l)}, \mathbf{S}^{(l)}{,\alpha^{(l)}})$\;
\BlankLine
\BlankLine
\Return $\mathbf{b}^*,\; \kappa^*,\; \mathbf{S}^*,\;{\alpha^*}$\;
\caption{Long-Timescale Optimization}
\label{algorithm:long_timescale}
\end{algorithm}

\subsubsection{AP Mode Selection}\label{sec:ap_mode}
We determine the AP mode vector $\mathbf{b}$ using the greedy search in Algorithm~\ref{algorithm:long_timescale}. Let $l$ denote the iteration index and $\mathbf{b}^{(l)}$ the AP mode vector at iteration $l$. The search is initialized from a communication-favorable configuration with $N^{(0)}=G-1$ transmit APs and $M^{(0)}=1$ receive AP. The initial receive AP is selected as the AP closest to the center of the coverage area. At each iteration, we examine all single-AP mode changes. Specifically, for each $g\in\{1,\ldots,G\}$, the $g$-th entry of $\mathbf{b}^{(l)}$ is flipped to generate the candidate AP mode vector $\mathbf{b}^{(l,g)}$. For each candidate, the sensing subcarrier allocation $(\kappa^{(l,g)},\mathbf{S}^{(l,g)})$ and power factor $\alpha^{(l,g)}$ are optimized using the procedures detailed later in Sections~\ref{sec:subcarrier_alloc} and~\ref{sec:pwoer_fact_opt}, respectively. Candidates that do not satisfy the constraint on the average sum-SE, $\widetilde{\mathrm{R}}_{\rm avg}(\mathbf{b}^{(l,g)},\mathbf{S}^{(l,g)},\alpha^{(l,g)})\geq \widetilde{\gamma}_{\rm avg}$, are discarded. Among the remaining candidates, the AP mode vector that minimizes the maximum image entropy across the synthetic scenarios is selected. The procedure terminates when no single-AP mode change yields further improvement.

\subsubsection{Subcarrier Allocation}\label{sec:subcarrier_alloc}

{For a given AP mode vector $\mathbf{b}$, with $N=\|\mathbf{b}\|_0$ transmit APs, problem \textbf{P2} allows a general sensing allocation in which different transmit APs can employ different number of sensing subcarriers, denoted by $\{\kappa_n\}_{n\in\mathcal{T}}$. To simplify the design, we consider a structured version of \textbf{P2} that imposes an equal allocation constraint,  whereby each transmitting AP is assigned the same number of sensing subcarriers $\kappa$. Consequently, the total number of sensing subcarriers becomes $S_{\rm sen} = \kappa N$, while the remaining $S_{\rm com} = S - \kappa N$ subcarriers are allocated for communication.}
{The sensing subcarriers are assigned to the transmit APs using a \emph{maximally interleaved} allocation. 
Let the $N$ transmit APs be indexed as $n \in \{1,\ldots,N\}$. The sensing subcarriers assigned to the $n$-th transmitter are given by}
\begin{equation}\label{eqn:interleaved}
{\mathcal{S}_{n,\rm sen} =
\{n,\; n+N,\; n+2N,\; \ldots,\; n+(\kappa-1)N\}.}
\end{equation}
Under this allocation, each transmit AP is assigned $\kappa$ uniformly spaced sensing subcarriers with stride $N$, spanning an effective bandwidth of $\kappa N\Delta_f$. The bistatic range resolution of each transmit-receive AP pair $(n,m)$ is inversely proportional to the effective bandwidth spanned by $\mathcal{S}_{n,\rm sen}$. Therefore, the interleaved allocation improves the range resolution by maximizing this effective bandwidth, even for small $\kappa$ values.

{ The feasible range of $\kappa$ is $\{1, \ldots, \kappa_{\rm max}\}$, where $\kappa_{\rm max}$ is the largest integer satisfying the constraint on the {average} sum-SE with fixed $\mathbf{b}$ and at $\alpha=1$}:
 \begin{align}\label{eqn:kappa_max}
 \kappa_{\rm max} = \max \big\{ \kappa \in \mathbb{Z}^+ :
 {\widetilde{\mathrm{R}}_{\rm avg}(\mathbf{b}, \mathbf{S}_\kappa,\alpha=1) \geq\widetilde{\gamma}_{\rm avg}} 
 \big\}.
 \end{align}
{For a given $\mathbf{b}$ and each feasible $\kappa \in \{1,\ldots,\kappa_{\rm max}\}$, the sensing subcarrier allocation matrix $\mathbf{S}_\kappa$ is constructed using~\eqref{eqn:interleaved}.}

\subsubsection{Power Factor Selection}\label{sec:pwoer_fact_opt}
{For the resulting pair $(\mathbf{b},\mathbf{S}_\kappa)$, the power allocation factor $\alpha$ is selected as the minimum value required to satisfy the {average} sum-SE constraint, i.e.,}
\begin{align}\label{eqn:opt_alpha_LTO}
     \alpha^*= \min \big\{\alpha \in [0,1] :
    {\widetilde{\mathrm{R}}_{\rm avg}(\mathbf{b},\mathbf{S}_\kappa,\alpha) \geq \widetilde{\gamma}_{\rm avg}} \big\}.\!
\end{align}
{Since {$\widetilde{\mathrm{R}}_{\rm avg}(\mathbf{b},\mathbf{S}_\kappa,\alpha)$} is monotonically increasing with $\alpha$, the optimal $\alpha^*$ can be obtained using bisection search over $[0,1]$.}

\subsubsection{Joint LTO Framework}
{The overall LTO framework jointly optimizes $\{\mathbf{b},\mathbf{S},\alpha\}$ to solve \textbf{P2}. The optimization is initialized from a communication-favorable operating point with: i) $N^{(0)}=G-1$ transmit APs and $M^{(0)}=1$ receive AP, where the receive AP is selected closest to the center of the coverage area; and ii) $\kappa^{(0)}=1$ subcarrier per transmit AP.  Under this initialization, only $G-1$ subcarriers are used for sensing, and $S-(G-1)$ subcarriers for communication. Since $S\gg G$, the bandwidth reduction relative to the communication-only reference is small. Therefore, this initialization provides a communication-favorable starting point for the greedy search. The complete LTO procedure is summarized in Algorithm~\ref{algorithm:long_timescale}.

\begin{algorithm}[t]
\DontPrintSemicolon
\SetAlgoLined
\KwIn{ $\{\mathbf{b},\mathbf{S}\}$ from LTO stage, current communication channels $\{\mathbf{G}[s]\}$, sum-SE budget $\beta$, AP positions $\{\mathbf{p}_g\}_{g=1}^G$, UE positions $\{\mathbf{u}_\ell\}_{\ell=1}^{L}$.}
\KwOut{Optimal $\alpha^*$}
\BlankLine
{Set $\gamma\gets\beta \mathrm{R}_{\max}$, where $\mathrm{R}_{\max}=\mathrm{R}(\mathbf{b}=\mathbf{1}_G,\kappa=0,\alpha=1)$ using current channels $\{\mathbf{G}[s]\}$\;}
\While{$\mathrm{R}(\mathbf{b},\mathbf{S}, \alpha=1) < \gamma$}{
    \If{$\kappa>1$}{
    {Set $\kappa\gets\kappa-1$}\;
        {Update $\mathcal{S}_{n,\rm sen}$ using \eqref{eqn:interleaved} $\forall n$}\;}

        \Else{
    Identify the weakest UE\; $\ell^* = \underset{\ell}{\arg\min}\,\, \mathrm{R}_\ell(\mathbf{b}, \mathbf{S},\alpha= 1)$\;
    Find the closest receive AP\; $g^* = \underset{g\in\mathcal{R}(\mathbf{b})}{\arg\min} \|\mathbf{p}_g - \mathbf{u}_{\ell^*}\|$\;
    Update $\mathbf{b}_{g^*} \gets 1$ \; 
    Update $\mathcal{S}_{g^*, \rm sen}$ using \eqref{eqn:interleaved} with current $\kappa$\;}}
Evaluate $\alpha^*$ using \eqref{eqn:alpha_opt}\;
\Return $\alpha^*$\;
\caption{Short-Timescale Optimization}
\label{algorithm:short_timescale}
\end{algorithm}

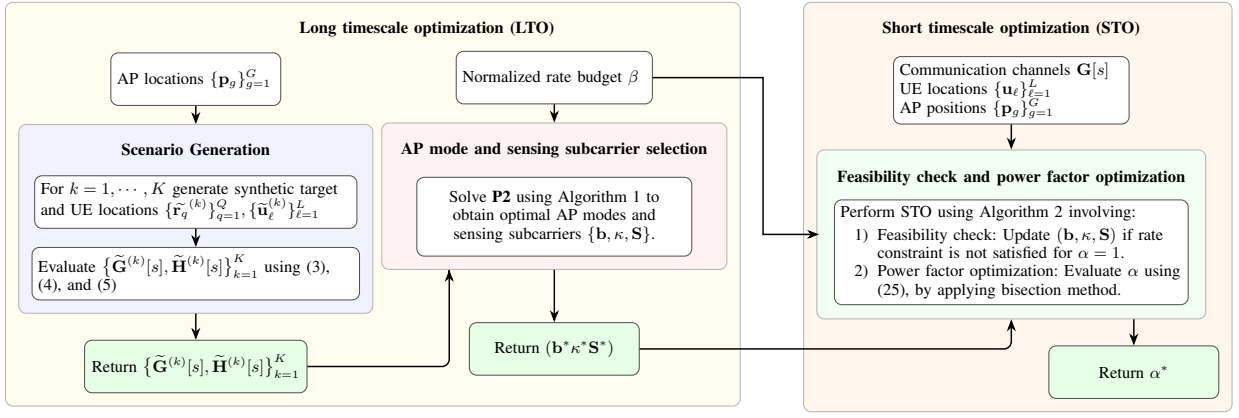
\begin{figure*}[t]
    \centering
    \begin{subfigure}[h]{\linewidth}
        \centering
        \resizebox{0.9\textwidth}{!}{\begin{tikzpicture}[
    font=\rmfamily\small,
    >=Stealth,
    node distance=1.2cm,
    base/.style={rectangle, rounded corners, draw=black!70, text centered, minimum height=2.8cm, fill=white},
    container/.style={rectangle, rounded corners, draw=black!30, fill=#1, inner sep=10pt},
    header/.style={font=\sffamily\bfseries, text depth=0.5ex},
    proc/.style={base, text width=2.8cm, fill=white},
    logic/.style={base, text width=3cm, fill=gray!10, dashed},
    io/.style={base, text width=2.5cm, fill=blue!5},
    var/.style={base, text width=4cm, fill=green!10, line width=1pt},
    legend/.style={circle, draw=black!50, minimum size=3mm, inner sep=0pt}
]

    \node (header_scenario) {\textbf{Scenario Generation}};
 \node (tar_ue_scenario) [proc, 
   below=  0.25cm of header_scenario, 
    text width=5.9cm, 
    minimum height=1cm, align=flush left] 
    { For $k=1,\cdots,K$ generate synthetic target and UE locations $\{\widetilde{\mathbf{r}_q}^{(k)}\}_{q=1}^{Q}, \{\widetilde{\mathbf{u}}_{\ell}^{(k)}\}_{\ell=1}^{L}$};

 \node (syn_channels) [proc,below=0.35cm of tar_ue_scenario, text width=6.0cm, minimum height=1cm, 
    align=flush left] {Evaluate $\big\{\widetilde{\mathbf{G}}^{(k)}[s],\widetilde{\mathbf{H}}^{(k)}[s]\big\}_{k=1}^K$ using \eqref{eq:comm_channel_mat_LOS}, \eqref{eq:comm_channel_mat_scat}, and \eqref{eq:csensing_channel_mat}
    };
    
    \begin{scope}[on background layer]
        \node (bg_scenario) [container=blue!5, fit=(header_scenario) (syn_channels) (tar_ue_scenario) , inner sep=7pt] {};
    \end{scope}
    
 \node (inputs) [proc, 
   above=  0.35cm of bg_scenario, 
    text width=3cm, 
    minimum height=1cm, align=flush left] 
    { AP locations $\{\mathbf{p}_g\}_{g=1}^{G}$};
\node (Return_scenario) [proc, below=0.5cm of bg_scenario, text width=4cm, minimum height=1cm, fill=green!10, align=center] { Return $\big\{\widetilde{\mathbf{G}}^{(k)}[s],\widetilde{\mathbf{H}}^{(k)}[s]\big\}_{k=1}^K$};

\draw [->, thick] (inputs) -- (bg_scenario);
\draw [->, thick] (tar_ue_scenario) -- (syn_channels);
\draw [->, thick] (bg_scenario) -- (Return_scenario);

\node (header_long) [right=2.25cm of header_scenario]  {\textbf{AP mode and sensing subcarrier selection}};
 \node (process_LTO) [proc, below=0.25cm of header_long, text width=5cm, minimum height=1.5cm, 
    align=center] {
    Solve \textbf{P2} using Algorithm \ref{algorithm:long_timescale} to obtain optimal AP modes and sensing subcarriers $\{\mathbf{b},\kappa,\mathbf{S}\}$.
    };    
\begin{scope}[on background layer]
    \node (bg_LTO) [container=red!5, fit= (header_long) (process_LTO) , inner sep=7.
    pt] {};
\end{scope}
\node (Return_LTO) [proc, below=1cm of bg_LTO, text width=3cm, minimum height=1cm, fill=green!10, align=center] { Return $(\mathbf{b}^{*} \kappa^{*} \mathbf{S}^{*})$};
\node (inputs_long) [proc, 
   above= 0.35cm of bg_LTO, 
    text width=3.5cm, 
    minimum height=1cm, align=flush left] 
     {Normalized rate budget $\beta$};
\node (title_LTO) [ above left=0.15cm and -2.0cm of inputs_long]  {\textbf{Long timescale optimization (LTO)}};

\draw [->, thick, rounded corners] (Return_scenario.east) 
    -| ($(bg_LTO.south) - (2cm, 0.15cm)$) 
    -- ($(bg_LTO.south) - (2cm, 0cm)$);
    
\draw [->, thick] (inputs_long) -- (bg_LTO);
\draw [->, thick] (bg_LTO) -- (Return_LTO);

\node (title_STO) [right=6cm of title_LTO]  {\textbf{Short timescale optimization (STO)}};
\node (inputs_short) [proc, 
    below= 0.25cm of title_STO, 
    text width=4.25cm, 
    minimum height=1cm, align=flush left] 
    {
    Communication channels $\mathbf{G}[s]$\\
    UE locations $\{\mathbf{u}_\ell\}_{\ell=1}^L$\\
    AP positions $\{\mathbf{p}_g\}_{g=1}^{G} $};
 \node (header_short) [below=0.75cm of inputs_short]  {\textbf{Feasibility check and power factor optimization}};
 \node (process_STO) [proc, below=0.15cm of header_short, text width=6.5cm, minimum height=1.5cm, 
    align=flush left] {
    Perform STO using Algorithm \ref{algorithm:short_timescale} involving:
    \begin{enumerate}
        \item Feasibility check: Update $(\mathbf{b},\kappa,\mathbf{S})$ if rate constraint is not satisfied for $\alpha=1$.
        \item Power factor optimization:  Evaluate $\alpha$ using \eqref{eqn:alpha_opt}, by applying bisection method.
    \end{enumerate}  
};    

\begin{scope}[on background layer]
    \node (bg_STO) [container=green!5, fit=(header_short) (process_STO) , inner sep=7.5pt] {};
\end{scope}   
\node (Return_STO) [proc, below right=0.5cm and -3cm of bg_STO, text width=3cm, minimum height=1cm, fill=green!10, align=center] { Return $\alpha^*$};

\draw [->, thick] (inputs_short) -- (bg_STO);
\draw [->, thick] (Return_STO.north |- bg_STO.south) -- (Return_STO.north);
\draw [->, thick, rounded corners] (Return_LTO.east) 
    -| ($(bg_STO.south) - (0, 0.25cm)$) 
    -- (bg_STO.south); 
\draw [->, thick] (inputs_long.east) -- ([xshift=2.1cm]inputs_long.east) |- ([xshift=0.05cm]bg_STO.west);

    \begin{scope}[on background layer]
        \node (bg_1)  [container=yellow!7, fit=(bg_scenario) (bg_LTO) (Return_LTO) (Return_scenario) (inputs) (inputs_long) (title_LTO), inner sep=7pt] { };
    \end{scope}
    \begin{scope}[on background layer]
        \node (bg_2)  [container=orange!7, fit= (bg_STO) (inputs_short) (Return_STO) (title_STO), inner sep=7pt] { };
    \end{scope}
    
 \begin{scope}[on background layer]
        \node (bg_scenario_x) [container=blue!5, fit=(header_scenario) (syn_channels) (tar_ue_scenario), inner sep=7pt] {};
    \end{scope}
\begin{scope}[on background layer]
    \node (bg_LTO_x) [container=red!5, fit= (header_long) (process_LTO) , inner sep=7.
    pt] {};
\end{scope}
\begin{scope}[on background layer]
    \node (bg_STO_x) [container=green!5, fit=(header_short) (process_STO) , inner sep=7.5pt] {};
\end{scope}
\end{tikzpicture}}
    \end{subfigure}
    \caption{Flowchart of the joint optimization framework, illustrating the interaction between long-timescale optimization (AP mode selection and subcarrier allocation), and short-timescale optimization (power-factor allocation).\vspace{-20pt}}
    \label{fig:flow_chart}    
\end{figure*}


\subsection{Short-Timescale Optimization (STO)}\label{sec:power_opt}
{The STO stage is initialized using the AP mode vector and subcarrier allocation matrix $(\mathbf{b},\mathbf{S})$ obtained from the LTO stage. The corresponding LTO power factor $\alpha$ is discarded. For a given communication budget $\beta$ and instantaneous channels $\mathbf{G}[s]$, the sum-SE threshold is computed as $\gamma=\beta \mathrm{R}_{\max}$, where $\mathrm{R}_{\max}=\mathrm{R}(\mathbf{b}=\mathbf{1}_G,\kappa=0,\alpha=1)$
denotes the maximum achievable sum-SE for the current channel realization, when all resources are allocated for communication.}

{Starting from the LTO-generated configuration $(\mathbf{b},\mathbf{S})$, the STO first checks the feasibility of the instantaneous sum-SE constraint under full communication power, i.e., $\mathrm{R}(\mathbf{b},\mathbf{S},\alpha=1)\geq \gamma$.
If the constraint is satisfied, the optimal $\alpha$ is set as the minimum value required to meet the sum-SE threshold:}
\begin{equation}\label{eqn:alpha_opt}
  \alpha^* = \min \big\{\alpha \in [0,1] :
    \mathrm{R}(\mathbf{b},\mathbf{S}, \alpha) \geq \gamma \big\}.
\end{equation}
{Since $\mathrm{R}(\mathbf{b},\mathbf{S},\alpha)$ is monotonically increasing with $\alpha$, the optimal $\alpha^*$ can be  determined using a bisection search over~$[0,1]$.}

{If the sum-SE constraint is violated even at $\alpha=1$, i.e., $\mathrm{R}(\mathbf{b},\mathbf{S},\alpha=1)<\gamma,$ 
the LTO-generated configuration becomes infeasible for the instantaneous channel realization. In this case, feasibility is restored by two sequential updates:}
\begin{enumerate}
    \item {First, the number of sensing subcarriers per transmit AP $\kappa$ is decremented in steps $\kappa \leftarrow \kappa - 1$,  provided $\kappa > 1$. The corresponding subcarrier allocation matrix $\mathbf{S}_\kappa$ is updated via \eqref{eqn:interleaved}. This step systematically releases $N$ sensing subcarriers for communication per iteration, while preserving the AP topology and spatial imaging diversity. The resulting $\kappa$ is chosen as the maximum value that restores feasibility at $\alpha=1$:}
\begin{align}\label{eqn:opt_kappa_STO}
\kappa^* = \max \big\{\kappa \geq 1 : \mathrm{R}(\mathbf{b}, \kappa, \alpha=1) \geq \gamma \big\}.
\end{align}
{If no such $\kappa$ exists, the algorithm sets $\kappa = 1$ and proceeds to the next step.}

\item {If the communication constraint remains infeasible after reducing $\kappa$ to its minimum value, $\kappa=1$, the AP mode vector $\mathbf{b}$ is updated iteratively by converting receive APs into transmit APs. At each iteration, the UE with the minimum SE is identified as}
\begin{align}\label{eq:weakest_UE}
\ell^* = \underset{\ell}{\arg\min}\,\mathrm{R}_{\ell}(\mathbf{b},\kappa=1,\alpha=1),
\end{align}
{where $\mathrm{R}_{\ell}(\mathbf{b},\kappa=1,\alpha)={1}/{S}\sum_{s\in\mathcal{S}_{\rm com}}\log_2(1+\mathrm{SINR}_{\ell,s})$.
The receive AP closest to this UE is then selected as}
\begin{equation}\label{eq:closest_rx_AP}
g^*=\underset{g\in\mathcal{R}(\mathbf{b})}{\arg\min}\,\|\mathbf{p}_g-\mathbf{u}_{\ell^*}\|,
\end{equation}
{and switched to transmit mode, i.e., $[\mathbf{b}]_{g^*}\leftarrow1$. The sensing subcarrier allocation is subsequently updated using \eqref{eqn:interleaved} with $\kappa=1$. This process continues until the communication constraint is satisfied.}
\end{enumerate}
Once the communication constraint is satisfied through updated $\kappa$ and $\mathbf{b}$, optimal $\alpha$ is determined using \eqref{eqn:alpha_opt}.  The complete STO procedure is summarized in Algorithm~\ref{algorithm:short_timescale}. The overall interaction between the offline LTO stage and the online STO updates is summarized in Fig.~\ref{fig:flow_chart}.

\vspace{-8pt}
\subsection{Convergence Analysis}
{We next analyze the convergence of the LTO and STO stages of the proposed resource allocation framework. }  

\subsubsection{Convergence of LTO}
Algorithm~\ref{algorithm:long_timescale} performs a greedy coordinate-descent search over the AP mode vector $\mathbf{b}$, while jointly optimizing $(\kappa,\mathbf{S})$ and $\alpha$ to minimize the maximum expected image entropy across $K$ scenarios. 
Algorithm~\ref{algorithm:long_timescale} is guaranteed to terminate at a local optimum of the restricted search problem induced by the interleaved equal-subcarrier allocation, as explained in the following paragraph.

{For $G$ APs, the AP mode vector $\mathbf{b}\in\{0,1\}^G$ admits at most $2^G$ configurations. For a fixed $\mathbf{b}$, the parameter $\kappa$ is bounded as $1\leq \kappa \leq \kappa_{\max}(\mathbf{b})$, where $\kappa_{\max}(\mathbf{b})$ is obtained using \eqref{eqn:kappa_max}. For a finite number of subcarriers $S$,  $\kappa_{\max}(\mathbf{b})$ is also finite. 
The sensing allocation matrix $\mathbf{S}$ is uniquely determined by $(\mathbf{b},\kappa)$ using \eqref{eqn:interleaved}. 
For fixed $(\mathbf{b},\mathbf{S})$, the power factor $\alpha$ is obtained from \eqref{eqn:opt_alpha_LTO}, using bisection search over $[0,1]$. Since $\widetilde{\mathrm{R}}_{\rm avg}(\mathbf{b},\mathbf{S},\alpha)$  is  continuous and non-decreasing function of $\alpha$, the bisection search converges in a finite number of steps. Hence, the overall feasible search~space~is~finite.} 
{Further, at iteration $l$, a candidate update is accepted only if: i) it satisfies the constraint on the {average sum-SE}, i.e., $\widetilde{\mathrm{R}}_{\rm avg}(\mathbf{b}^{(l,g)},\kappa^{(l,g)}\!=\!1,\alpha\!=\!1)\!\geq\!\widetilde{\gamma}_{\rm avg}$; and ii) it strictly decreases the maximum expected entropy objective, i.e.,
$F^{(l+1)} < F^{(l)}.$ Hence, the sequence $\{F^{(l)}\}$ is monotonically decreasing and lower bounded by zero. The strict descent criterion also prevents revisiting previously explored configurations, avoiding cyclic updates. Since the feasible set is finite, Algorithm~\ref{algorithm:long_timescale} is guaranteed to terminate after a finite number of iterations at~a~local~optimum~\cite{Conv}.}

\subsubsection{Convergence of STO}
The STO stage consists of a sum-SE feasibility check, followed by a bisection search over $\alpha$. The convergence of the two stages are analyzed~separately.
\paragraph{Termination of the Feasibility Check Stage}
The feasibility check stage is designed to recover communication feasibility when the LTO configuration does not satisfy the instantaneous sum-SE constraint.
It consists of two sequential steps. In the first step, the number of sensing subcarriers per transmit AP $\kappa$ is iteratively reduced as $\kappa\gets\kappa-1$ until the sum-SE constraint is satisfied or the lower bound $\kappa=1$ is reached. Since $\kappa$ is integer-valued, bounded, and strictly decreases at each iteration, this stage terminates in a finite number of steps. 
If the sum-SE constraint remains violated at $\kappa=1$, the algorithm proceeds to the second step. At each iteration, one receive AP is  switched to transmit mode via $[\mathbf{b}]_{g^*}\leftarrow 1$, where $g^*$ is obtained using \eqref{eq:closest_rx_AP}. Since $\mathbf{b}\in\{0,1\}^G$ admits $2^G$ configurations, and the updates are unidirectional, previously visited states cannot reoccur. Consequently, the AP mode update stage also terminates in a finite number of iterations. 
Therefore, both stages operate over finite search spaces with monotonic updates, guaranteeing finite-time termination.

\paragraph{Convergence of the Bisection Search}
Once feasible $(\mathbf{b}, \mathbf{S})$ is obtained, the power factor $\alpha$ is computed using bisection over $[0,1]$. For fixed $(\mathbf{b}, \mathbf{S})$, the sum-SE $\mathrm{R}(\mathbf{b},\mathbf{S},\alpha)$ is continuous and non-decreasing in $\alpha$. Feasibility at $\alpha=1$, i.e., $\mathrm{R}(\mathbf{b},\mathbf{S},\alpha=1)\geq \gamma$, ensures at least one solution $\alpha^* \in [0,1]$ satisfying $\mathrm{R}(\mathbf{b},\mathbf{S},\alpha^*) \geq \gamma$ exists. 
We select the smallest feasible power factor,
$\alpha^*=\inf\left\{ \alpha\in[0,1]:\mathrm{R}(\mathbf{b},\mathbf{S},\alpha)\geq\gamma\right\}.$ 
The bisection method converges to $\alpha^*$ by iteratively shrinking the interval $[0,1]$ based on feasibility checks. This yields convergence after a finite number of iterations.

\vspace{-2pt}
\section{Numerical Results}\vspace{-2pt}\label{sec:Num_results}
In this section, we evaluate the communication and imaging performance of the proposed framework and investigate the resulting sensing-communication trade-offs.

\vspace{-8pt}
\subsection{Simulation Setup and Performance Metrics}
We consider a network of $G=100$ single-antenna APs deployed on a regular grid over a $10\,\text{m} \times 10\,\text{m}$ area. We assume $L=5$ UEs and $Q=3$ targets present in the coverage area. The system operates at $f_0=1$ GHz, with $S=1000$ subcarriers and subcarrier spacing $\Delta_f=120$ kHz, resulting in a total bandwidth of $W=120$ MHz. The total transmit power $P$ is set to $10$ dBm, and the precoder $\mathbf{F}_{\rm com}[s]$ is designed based on maximal-ratio (MR) principle.  The noise power at both the receive APs and the UEs is given by $\sigma_n^2 = N_0\Delta_f$, where $N_0=-174$~dBm/Hz is the noise power spectral density~\cite{Bjornson2020Scalable}.
For the LTO stage, we use one representative synthetic scenario to keep the offline entropy-based search computationally tractable. The LTO configuration is tested over multiple independent STO realizations with randomly generated UE~and~target~locations.

We next introduce the performance metrics used in the numerical evaluation. For communication, we use the sum-SE defined in Section~\ref{subsec:sumrate}. For sensing, we consider both image quality and target estimation accuracy. Image quality is evaluated using the entropy metric defined in Section~\ref{subsec:entropy}, while target detection and localization performance are evaluated using the generalized optimal sub-pattern assignment (GOSPA) metric~\cite{GOSPA}. GOSPA jointly accounts for localization error, missed detections, and false alarms. 

\begin{figure}[t]
    \centering
    \begin{subfigure}[b]{0.75\linewidth}
        \centering
        \includegraphics[width=\linewidth,height=0.9\linewidth]{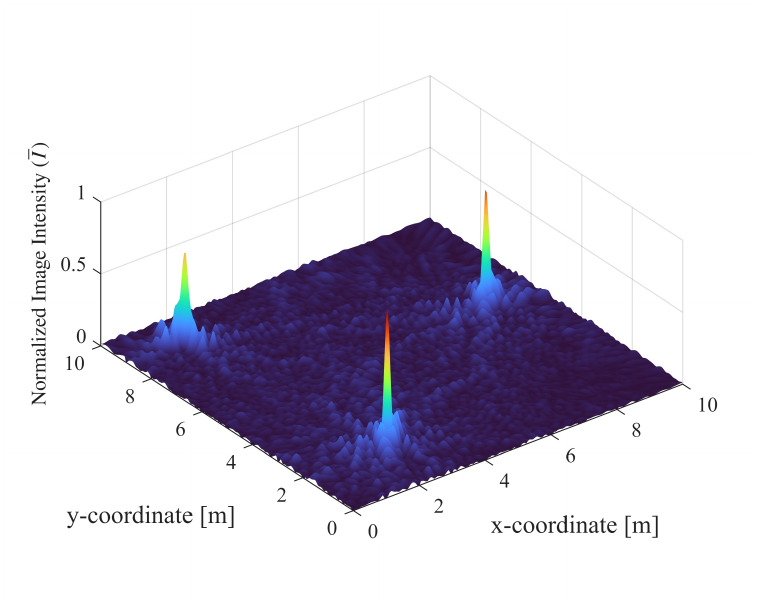}\vspace{-22pt}
        \caption{}
        \label{fig:Im_opt}
            \vspace{-1pt}
    \end{subfigure}
    \begin{subfigure}[b]{0.75\linewidth}
        \centering
         \includegraphics[width=\linewidth]{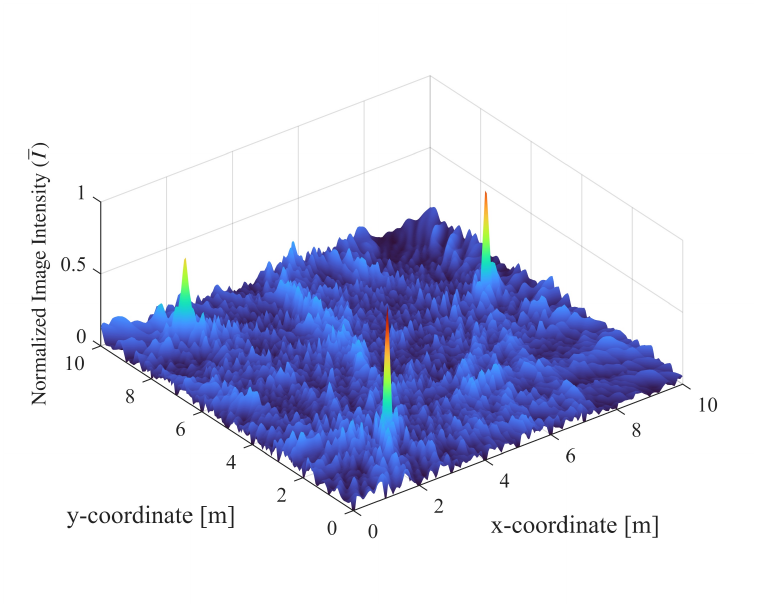}\vspace{-20pt}
        \caption{}
        \label{fig:Im_unopt}
    \end{subfigure}
        \vspace{-10pt}
  \caption{Normalized reconstructed image intensity for a representative STO realization for (a) $\beta=0.25$ and (b) $\beta=0.9$, where $\beta$ denotes the required fraction of the maximum communication sum-SE.}
    \label{fig:Images}
\end{figure}

\vspace{-8pt}
\subsection{Imaging Performance Under Different Sum-SE Budgets}\vspace{-3pt}

We begin by visualizing the normalized image intensities in Fig.~\ref{fig:Images} for one representative STO realization under two sum-SE budgets, $\beta=\{0.25,0.9\}$. In both cases, the true target locations appear as dominant peaks. However, for $\beta=0.25$, the sidelobe levels are significantly lower, resulting in an image entropy of $9.43$ bits. In contrast, for $\beta=0.9$, the entropy increases to $11.69$ bits, and produces more pronounced secondary peaks. Fig.~\ref{fig:Ent_cdf} and Fig.~\ref{fig:GOSPA_cdf} show the CDFs of the image entropy and GOSPA metric, respectively, over $100$ STO realizations with random UE and target locations. As $\beta$ increases, both CDFs shift toward higher values, indicating degraded imaging and reduced target detection and localization performance. Specifically, for $\beta=\{0.25,0.75,0.9\}$, approximately $\{93,67,53\}\%$ of the realizations achieve $\mathrm{GOSPA}<10^{-3}$, corresponding to correct detection and accurate localization of all targets, without missed detections or false alarms. This degradation arises because a larger $\beta$ allocates more power and subcarriers to communication and requires more APs to operate in transmit mode, reducing both the sensing resources and the flexibility of AP mode selection.
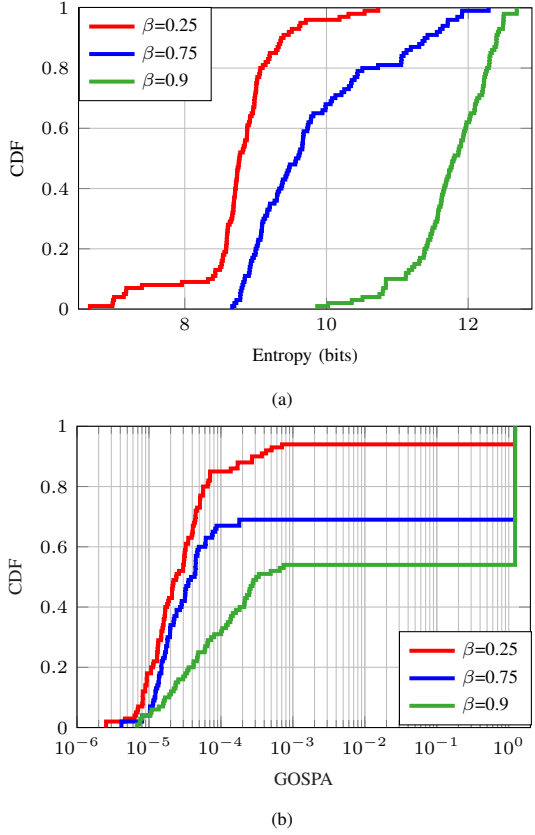
\begin{figure}[t]
     \begin{subfigure}[b]{0.75\linewidth}\hspace{-15pt}
        \centering
%
\definecolor{mycolor1}{rgb}{0.23100,0.66600,0.19600}%
\definecolor{mycolor2}{rgb}{0.12941,0.12941,0.12941}%
\begin{tikzpicture}

\begin{axis}[%
width=0.903\columnwidth,
height=0.6\columnwidth,
at={(0\columnwidth,0\columnwidth)},
scale only axis,
xmin=6.5,
xmax=12.9,
xlabel style={font=\color{mycolor2},font=\scriptsize},
xlabel={Entropy (bits)},
ymin=0,
ymax=1,
ylabel style={font=\color{mycolor2},font=\scriptsize},
ylabel={CDF},
axis background/.style={fill=white},
xmajorgrids,
ymajorgrids,
tick label style={font=\scriptsize},
legend style={at={(0.00042,0.69)}, anchor=south west, legend cell align=left, align=left},font=\scriptsize
]
\addplot[const plot, color=red, line width=1.5pt] table[row sep=crcr] {%
6.660801568	0\\
6.660801568	0.01\\
6.979343406	0.02\\
6.986639669	0.03\\
7.001253773	0.04\\
7.139423234	0.05\\
7.169504607	0.06\\
7.174594863	0.07\\
7.39471145	0.08\\
7.959651393	0.09\\
8.328519988	0.1\\
8.396229594	0.11\\
8.431092164	0.12\\
8.431212653	0.13\\
8.497676068	0.14\\
8.518112023	0.15\\
8.521659697	0.16\\
8.536954992	0.17\\
8.538367323	0.18\\
8.557804528	0.19\\
8.583265888	0.2\\
8.584137423	0.21\\
8.589102918	0.22\\
8.592653153	0.23\\
8.593375498	0.24\\
8.596415665	0.25\\
8.602280916	0.26\\
8.605898239	0.27\\
8.609371063	0.28\\
8.633844728	0.29\\
8.65567186	0.3\\
8.670086174	0.31\\
8.671398782	0.32\\
8.684993458	0.33\\
8.686476547	0.34\\
8.688473586	0.35\\
8.693583198	0.36\\
8.694983946	0.37\\
8.70603636	0.38\\
8.708283802	0.39\\
8.712477569	0.4\\
8.719899095	0.41\\
8.720693893	0.42\\
8.728601915	0.43\\
8.73523409	0.44\\
8.738632721	0.45\\
8.740180365	0.46\\
8.745837391	0.47\\
8.75975685	0.48\\
8.760695473	0.49\\
8.769165985	0.5\\
8.773422335	0.51\\
8.782244179	0.52\\
8.817816663	0.53\\
8.829914768	0.54\\
8.845401367	0.55\\
8.849705964	0.56\\
8.875367415	0.57\\
8.877315281	0.58\\
8.879465682	0.59\\
8.880686227	0.6\\
8.882779249	0.61\\
8.904508881	0.62\\
8.917989451	0.63\\
8.920191503	0.64\\
8.943412996	0.65\\
8.96253002	0.66\\
8.965320793	0.67\\
8.981168538	0.68\\
8.982720796	0.69\\
8.983628827	0.7\\
8.990194363	0.71\\
8.996606674	0.72\\
9.007489768	0.73\\
9.007761252	0.74\\
9.008834451	0.75\\
9.017286	0.76\\
9.023203734	0.77\\
9.050832746	0.78\\
9.058750273	0.79\\
9.064702694	0.8\\
9.100773052	0.81\\
9.140165165	0.82\\
9.174440874	0.83\\
9.193977727	0.84\\
9.200050941	0.85\\
9.286925372	0.86\\
9.310310099	0.87\\
9.327668618	0.88\\
9.347518479	0.89\\
9.35953739	0.9\\
9.394244122	0.91\\
9.467532022	0.92\\
9.511096784	0.93\\
9.614329467	0.94\\
9.632140269	0.95\\
9.709170593	0.96\\
10.18443238	0.97\\
10.31032994	0.98\\
10.53792348	0.99\\
10.73075555	1\\
};
\addlegendentry{$\beta\text{=0.25}$}

\addplot[const plot, color=blue, line width=1.5pt] table[row sep=crcr] {%
8.668843507	0\\
8.668843507	0.01\\
8.699718151	0.02\\
8.728048911	0.03\\
8.782645308	0.04\\
8.783846136	0.05\\
8.788018235	0.06\\
8.800403129	0.07\\
8.81304315	0.08\\
8.825081869	0.09\\
8.842917589	0.1\\
8.849047786	0.11\\
8.90466693	0.12\\
8.915197533	0.13\\
8.917217593	0.14\\
8.928372545	0.15\\
8.933960779	0.16\\
8.953264755	0.17\\
8.975558776	0.18\\
8.999073037	0.19\\
8.999442123	0.2\\
9.014564264	0.21\\
9.030531939	0.22\\
9.036468421	0.23\\
9.069357861	0.24\\
9.077089314	0.25\\
9.080018033	0.26\\
9.088452787	0.27\\
9.089481245	0.28\\
9.094946708	0.29\\
9.111493251	0.3\\
9.134601313	0.31\\
9.162246678	0.32\\
9.166221328	0.33\\
9.195743679	0.34\\
9.202844814	0.35\\
9.267017101	0.36\\
9.299970728	0.37\\
9.302357395	0.38\\
9.3118613	0.39\\
9.332227136	0.4\\
9.364391419	0.41\\
9.375701946	0.42\\
9.383649766	0.43\\
9.418913454	0.44\\
9.435645029	0.45\\
9.450843471	0.46\\
9.470352838	0.47\\
9.482332711	0.48\\
9.559260946	0.49\\
9.568645986	0.5\\
9.603616721	0.51\\
9.617751703	0.52\\
9.627967201	0.53\\
9.658826167	0.54\\
9.660015606	0.55\\
9.665961772	0.56\\
9.668299025	0.57\\
9.675062825	0.58\\
9.677061982	0.59\\
9.72948208	0.6\\
9.735115763	0.61\\
9.746088363	0.62\\
9.779002047	0.63\\
9.783945624	0.64\\
9.815061721	0.65\\
9.943118524	0.66\\
9.98589163	0.67\\
9.987378184	0.68\\
10.0588605	0.69\\
10.07839536	0.7\\
10.12565593	0.71\\
10.21145799	0.72\\
10.25017817	0.73\\
10.31530201	0.74\\
10.34621113	0.75\\
10.35640207	0.76\\
10.39733406	0.77\\
10.43505942	0.78\\
10.439516	0.79\\
10.50549256	0.8\\
10.72731541	0.81\\
11.05372817	0.82\\
11.05839901	0.83\\
11.05887287	0.84\\
11.08841062	0.85\\
11.13027572	0.86\\
11.19116486	0.87\\
11.30830166	0.88\\
11.35452152	0.89\\
11.38452607	0.9\\
11.42565812	0.91\\
11.55776204	0.92\\
11.60341506	0.93\\
11.63968487	0.94\\
11.71043323	0.95\\
11.71311664	0.96\\
11.84667186	0.97\\
11.91271846	0.98\\
11.91501016	0.99\\
12.28962985	1\\
};
\addlegendentry{$\beta\text{=0.75}$}

\addplot[const plot, color=mycolor1, line width=1.5pt] table[row sep=crcr] {%
9.869348475	0\\
9.869348475	0.01\\
10.02292096	0.02\\
10.35700052	0.03\\
10.50881343	0.04\\
10.74717208	0.05\\
10.79338466	0.06\\
10.80026697	0.07\\
10.83744107	0.08\\
10.8387006	0.09\\
10.83876846	0.1\\
11.11909607	0.11\\
11.12200503	0.12\\
11.16083021	0.13\\
11.20933676	0.14\\
11.24066075	0.15\\
11.30891372	0.16\\
11.31846334	0.17\\
11.37268468	0.18\\
11.38251968	0.19\\
11.38520403	0.2\\
11.41025744	0.21\\
11.42635415	0.22\\
11.43514479	0.23\\
11.46599843	0.24\\
11.46925391	0.25\\
11.48999356	0.26\\
11.49718874	0.27\\
11.52634927	0.28\\
11.53847264	0.29\\
11.56278396	0.3\\
11.56937822	0.31\\
11.57411311	0.32\\
11.57867795	0.33\\
11.59165684	0.34\\
11.61201707	0.35\\
11.62060148	0.36\\
11.62514139	0.37\\
11.62978643	0.38\\
11.64624551	0.39\\
11.65629209	0.4\\
11.66577405	0.41\\
11.68374208	0.42\\
11.69220125	0.43\\
11.70794377	0.44\\
11.72752894	0.45\\
11.7293406	0.46\\
11.75067573	0.47\\
11.76779122	0.48\\
11.77293088	0.49\\
11.79322924	0.5\\
11.80824392	0.51\\
11.86087812	0.52\\
11.86514898	0.53\\
11.87593492	0.54\\
11.89793542	0.55\\
11.90109507	0.56\\
11.92598657	0.57\\
11.93284181	0.58\\
11.94047513	0.59\\
11.95842034	0.6\\
11.96872602	0.61\\
11.97622313	0.62\\
12.01481661	0.63\\
12.02302183	0.64\\
12.03511444	0.65\\
12.08583369	0.66\\
12.09397518	0.67\\
12.09520827	0.68\\
12.1121774	0.69\\
12.12698984	0.7\\
12.15722346	0.71\\
12.19726574	0.72\\
12.20546164	0.73\\
12.20809864	0.74\\
12.22127343	0.75\\
12.22194182	0.76\\
12.22724941	0.77\\
12.23998739	0.78\\
12.25008476	0.79\\
12.26531766	0.8\\
12.29163271	0.81\\
12.29927242	0.82\\
12.30162384	0.83\\
12.30256255	0.84\\
12.31428007	0.85\\
12.33504228	0.86\\
12.3698719	0.87\\
12.39140302	0.88\\
12.39949082	0.89\\
12.39954978	0.9\\
12.40088525	0.91\\
12.41967203	0.92\\
12.43710387	0.93\\
12.49108375	0.94\\
12.49346642	0.95\\
12.49845961	0.96\\
12.49926642	0.97\\
12.50263707	0.98\\
12.68573402	0.99\\
12.68748021	1\\
};
\addlegendentry{$\beta\text{=0.9}$}

\end{axis}

\end{tikzpicture}%
        \caption{}
        \label{fig:Ent_cdf}
    \end{subfigure}
         \begin{subfigure}[b]{0.75\linewidth}\hspace{-15pt}
        \centering
%
\definecolor{mycolor1}{rgb}{0.23137,0.66667,0.19608}%
\definecolor{mycolor2}{rgb}{0.12941,0.12941,0.12941}%
\begin{tikzpicture}

\begin{axis}[%
width=0.903\columnwidth,
height=0.6\columnwidth,
at={(0\columnwidth,0\columnwidth)},
scale only axis,
xmode=log,
xmin=1e-06,
xmax=2,
xminorticks=true,
xlabel style={font=\color{mycolor2}\scriptsize},
xlabel={GOSPA},
ymin=0,
ymax=1,
ylabel style={font=\color{mycolor2}\scriptsize},
ylabel={CDF},
axis background/.style={fill=white},
xmajorgrids,
xminorgrids,
ymajorgrids,
tick label style={font=\scriptsize},
legend style={at={(0.71,0.0071)}, anchor=south west, legend cell align=left, align=left, font=\scriptsize}
]
\addplot[const plot, color=red, line width=1.5pt] table[row sep=crcr] {%
2.537028867e-06	0\\
2.537028867e-06	0.02\\
4.62621702e-06	0.03\\
6.319162023e-06	0.04\\
6.60312336e-06	0.05\\
6.95386255e-06	0.06\\
7.077491658e-06	0.07\\
8.103174568e-06	0.08\\
8.191003744e-06	0.1\\
8.209993038e-06	0.12\\
8.894794898e-06	0.14\\
9.296392463e-06	0.16\\
9.356306024e-06	0.18\\
1.071888854e-05	0.2\\
1.140768566e-05	0.21\\
1.16453163e-05	0.22\\
1.293706835e-05	0.24\\
1.303602341e-05	0.25\\
1.322243024e-05	0.26\\
1.324043725e-05	0.27\\
1.344951021e-05	0.28\\
1.348555574e-05	0.29\\
1.474643003e-05	0.3\\
1.48675102e-05	0.32\\
1.538735181e-05	0.33\\
1.563652299e-05	0.34\\
1.603586482e-05	0.36\\
1.644821047e-05	0.37\\
1.668628219e-05	0.38\\
1.676719442e-05	0.39\\
1.688863204e-05	0.4\\
1.784912925e-05	0.41\\
1.848688959e-05	0.42\\
1.895319032e-05	0.43\\
2.092757966e-05	0.46\\
2.127344401e-05	0.47\\
2.170710678e-05	0.48\\
2.193124184e-05	0.49\\
2.386925108e-05	0.5\\
2.420151321e-05	0.51\\
2.619880407e-05	0.52\\
2.987380709e-05	0.54\\
3.051108515e-05	0.55\\
3.074060223e-05	0.56\\
3.10939779e-05	0.57\\
3.134182143e-05	0.58\\
3.158695047e-05	0.59\\
3.246892521e-05	0.61\\
3.513617822e-05	0.63\\
3.931293693e-05	0.64\\
4.021435281e-05	0.65\\
4.06861482e-05	0.67\\
4.218015114e-05	0.68\\
4.373782678e-05	0.69\\
4.437961078e-05	0.7\\
4.459620567e-05	0.72\\
4.67219691e-05	0.73\\
5.115144972e-05	0.76\\
5.172773378e-05	0.77\\
5.626593232e-05	0.78\\
5.652060731e-05	0.8\\
6.563414037e-05	0.81\\
6.707106781e-05	0.82\\
7.07002157e-05	0.84\\
7.071067812e-05	0.85\\
0.0001370549879	0.86\\
0.0001707106781	0.88\\
0.0002707106781	0.89\\
0.0002707106781	0.9\\
0.0003707106785	0.91\\
0.0004247071068	0.92\\
0.0005071067815	0.93\\
0.0007071067817	0.94\\
1.2247	1\\
};
\addlegendentry{$\beta\text{=0.25}$}

\addplot[const plot, color=blue, line width=1.5pt] table[row sep=crcr] {%
4.149265216e-06	0\\
4.149265216e-06	0.02\\
8.043603044e-06	0.04\\
1.029672022e-05	0.05\\
1.032927287e-05	0.07\\
1.132051342e-05	0.08\\
1.148954412e-05	0.09\\
1.194567308e-05	0.1\\
1.227752428e-05	0.12\\
1.249863724e-05	0.13\\
1.303602341e-05	0.14\\
1.353647114e-05	0.15\\
1.373517563e-05	0.17\\
1.403165614e-05	0.18\\
1.488858095e-05	0.2\\
1.49530869e-05	0.22\\
1.561901023e-05	0.23\\
1.652830552e-05	0.24\\
1.671250234e-05	0.25\\
1.72904732e-05	0.27\\
1.800316663e-05	0.29\\
1.816431677e-05	0.3\\
1.961037225e-05	0.31\\
1.965612372e-05	0.32\\
1.970396603e-05	0.33\\
1.990901248e-05	0.34\\
2.196755413e-05	0.35\\
2.211571914e-05	0.36\\
2.261005416e-05	0.37\\
2.420151321e-05	0.39\\
2.698917715e-05	0.4\\
2.839912622e-05	0.41\\
2.854231959e-05	0.42\\
3.157917089e-05	0.43\\
3.177271936e-05	0.44\\
3.205018109e-05	0.45\\
3.205778608e-05	0.46\\
3.297019559e-05	0.47\\
3.521632085e-05	0.48\\
3.542200339e-05	0.49\\
3.839952793e-05	0.5\\
4.322063063e-05	0.51\\
4.423632302e-05	0.53\\
4.461809459e-05	0.55\\
4.47910899e-05	0.57\\
4.709657221e-05	0.58\\
4.713312408e-05	0.59\\
4.932837569e-05	0.6\\
6.072575893e-05	0.61\\
6.147976351e-05	0.63\\
7.567303583e-05	0.64\\
7.694201017e-05	0.65\\
8.333006159e-05	0.66\\
8.695977545e-05	0.67\\
0.0001800974893	0.69\\
1.2247	0.95\\
1.2247	0.96\\
1.2247	0.97\\
1.2247	0.99\\
1.2247	1\\
};
\addlegendentry{$\beta\text{=0.75}$}

\addplot[const plot, color=mycolor1, line width=1.5pt] table[row sep=crcr] {%
6.791311338e-06	0\\
6.791311338e-06	0.01\\
7.411065595e-06	0.02\\
7.604970914e-06	0.03\\
8.110448699e-06	0.04\\
1.051915869e-05	0.05\\
1.087323689e-05	0.06\\
1.371883318e-05	0.07\\
1.572322033e-05	0.08\\
1.604585995e-05	0.09\\
1.670115944e-05	0.1\\
1.9192648e-05	0.11\\
2.025090196e-05	0.12\\
2.211309278e-05	0.13\\
2.335252451e-05	0.14\\
2.354827128e-05	0.15\\
2.555048282e-05	0.16\\
2.99633076e-05	0.17\\
3.189332323e-05	0.18\\
3.392262931e-05	0.19\\
3.554308811e-05	0.2\\
4.163355014e-05	0.21\\
4.170894728e-05	0.22\\
4.65161182e-05	0.23\\
4.744083452e-05	0.24\\
4.836560533e-05	0.25\\
5.958230121e-05	0.26\\
6.057545437e-05	0.27\\
6.502515135e-05	0.28\\
6.584499273e-05	0.29\\
7.311135153e-05	0.3\\
8.066044782e-05	0.31\\
0.0001007123508	0.32\\
0.0001040257949	0.33\\
0.0001179371753	0.34\\
0.0001245130281	0.35\\
0.0001352286978	0.36\\
0.0001379082135	0.37\\
0.0001464464437	0.38\\
0.0001616070197	0.39\\
0.0001819520608	0.4\\
0.0002107570168	0.41\\
0.0002136816179	0.42\\
0.0002168176866	0.43\\
0.0002296906588	0.44\\
0.0002473514656	0.45\\
0.0002499341662	0.46\\
0.0002590219025	0.47\\
0.0002759811782	0.48\\
0.0002828188156	0.49\\
0.0003093223907	0.5\\
0.0003366944145	0.51\\
0.0004914223462	0.52\\
0.000657678938	0.53\\
0.0007453258089	0.54\\
1.2247	0.9\\
1.2247	0.91\\
1.2247	0.92\\
1.2247	0.93\\
1.2247	0.95\\
1.2247	0.96\\
1.2247	0.97\\
1.2247	0.99\\
1.224744871	1\\
};
\addlegendentry{$\beta\text{=0.9}$}

\end{axis}

\end{tikzpicture}%
        \caption{}
        \label{fig:GOSPA_cdf}
    \end{subfigure}\vspace{-5pt}
  \caption{CDF of the (a) reconstructed-image  entropy  and (b) GOSPA over $100$ STO realizations for different normalized communication budgets $\beta$.}
    \label{fig:Ent_GOSPAcdf}
\end{figure}

\begin{figure}[t]
    \centering
     \begin{subfigure}[h]{0.75\linewidth}
        \centering
%
\definecolor{mycolor1}{rgb}{0.06667,0.44314,0.74510}%
\definecolor{mycolor2}{rgb}{0.12941,0.12941,0.12941}%
\begin{tikzpicture}

\begin{axis}[%
width=0.903\columnwidth,
height=0.6\columnwidth,
at={(0\columnwidth,0\columnwidth)},
scale only axis,
xmin=0,
xmax=23,
xlabel style={font=\color{mycolor2},font=\scriptsize},
xlabel={Average sum-SE  (bits/sec/Hz)},
ymin=7.5,
ymax=13.5,
ylabel style={font=\color{mycolor2},font=\scriptsize},
ylabel={Average Entropy (bits)},
axis background/.style={fill=white},
xmajorgrids,
ymajorgrids,
tick label style={font=\scriptsize},
legend style={at={(0.00,0.75000)}, anchor=south west, legend cell align=left, align=left, font=\scriptsize, fill opacity=0.6, draw opacity=1, text opacity=1,
    legend columns=2}
]
\addplot [color=mycolor1, line width=1.5pt, mark=diamond, mark options={solid, mycolor1}]
  table[row sep=crcr]{%
0	8.8812\\
1.523	8.90232\\
3.5575	9.2257\\
5.2651	9.4562\\
7.1149	9.8229\\
9.2494	10.2299\\
10.6724	10.5752\\
11.38	10.94635\\
13.81	11.307261\\
14.0212	11.717261\\
};
\addlegendentry{Baseline}

\addplot [color=mycolor1, dashed, line width=1.5pt]
  table[row sep=crcr]{%
14.5	7\\
14.5	13.5\\
};
\addlegendentry{$\overline{\mathrm{R}}_{\max}^{\rm BL}$}

\addplot [color=red, line width=1.5pt, mark=triangle, mark options={solid, red}]
  table[row sep=crcr]{%
0	8.4575\\
3.546512	8.4944\\
6.408410469	8.5801009\\
9.56123	8.90789\\
12.13463564	9.12534729\\
15.51232	9.8328\\
17.60281881	10.12534729\\
18.6521	10.3981\\
20.0482017	10.54\\
21.5321	11.654\\
};
\addlegendentry{Proposed }

\addplot [color=red, dashed, line width=1.5pt]
  table[row sep=crcr]{%
22	7\\
22	13.5\\
};
\addlegendentry{$\overline{\mathrm{R}}_{\max}$}

\end{axis}

\end{tikzpicture}\vspace{-20pt}
    \caption{}
     \label{fig:EntvsSE_APdep}
    \end{subfigure}\vspace{-10pt}
    \begin{subfigure}[h]{0.75\linewidth}
        \centering
%
\definecolor{mycolor1}{rgb}{0.06667,0.44314,0.74510}%
\definecolor{mycolor2}{rgb}{0.12941,0.12941,0.12941}%
\begin{tikzpicture}

\begin{axis}[%
width=0.903\columnwidth,
height=0.6\columnwidth,
at={(0\columnwidth,0\columnwidth)},
scale only axis,
xmode=log,
xmin=1e-06,
xmax=1.000000001,
xminorticks=true,
xlabel style={font=\color{mycolor2}, font=\scriptsize},
xlabel={GOSPA},
ymin=0,
ymax=1,
ylabel style={font=\color{mycolor2}, font=\scriptsize},
ylabel={CDF},
axis background/.style={fill=white},
xmajorgrids,
xminorgrids,
ymajorgrids,
tick label style={font=\scriptsize},
legend style={at={(0.00179,0.778063)}, anchor=south west, legend cell align=left, align=left, font=\scriptsize}
]

\addplot[const plot, color=red, line width=1.5pt] table[row sep=crcr] {%
3.22349505e-06	0\\
3.22349505e-06	0.01\\
3.227616839e-06	0.02\\
7.297711782e-06	0.04\\
8.042206057e-06	0.05\\
8.314285323e-06	0.07\\
8.450342008e-06	0.08\\
9.211811831e-06	0.09\\
1.05591633e-05	0.1\\
1.19125871e-05	0.11\\
1.206771134e-05	0.12\\
1.278483941e-05	0.13\\
1.356665652e-05	0.14\\
1.691814848e-05	0.15\\
1.762980643e-05	0.16\\
2.161630216e-05	0.17\\
2.21534807e-05	0.18\\
2.255713267e-05	0.19\\
2.265611892e-05	0.2\\
2.414281846e-05	0.21\\
2.635554909e-05	0.22\\
2.643025025e-05	0.23\\
2.741974537e-05	0.25\\
2.805875066e-05	0.26\\
2.888111242e-05	0.27\\
2.933321225e-05	0.28\\
3.056348516e-05	0.29\\
3.307371612e-05	0.31\\
3.372063262e-05	0.32\\
3.45682835e-05	0.33\\
3.583150535e-05	0.34\\
3.603543791e-05	0.35\\
3.610559673e-05	0.36\\
3.882215839e-05	0.37\\
3.907826637e-05	0.38\\
3.935180812e-05	0.39\\
4.250331792e-05	0.4\\
4.2813525e-05	0.41\\
4.285162651e-05	0.42\\
4.341724317e-05	0.43\\
4.396111856e-05	0.44\\
5.026355899e-05	0.45\\
5.040336441e-05	0.46\\
5.204700503e-05	0.49\\
5.350067844e-05	0.5\\
5.359422803e-05	0.52\\
5.514361973e-05	0.53\\
5.6784347e-05	0.55\\
6.052410115e-05	0.56\\
6.145877032e-05	0.57\\
6.180869603e-05	0.59\\
6.239452641e-05	0.6\\
6.241813353e-05	0.61\\
7.109910286e-05	0.62\\
7.632040234e-05	0.63\\
7.786973935e-05	0.64\\
8.36541538e-05	0.65\\
8.376212077e-05	0.66\\
8.738805592e-05	0.67\\
9.257583818e-05	0.68\\
0.0001094948643	0.69\\
0.0001114324411	0.7\\
0.0001222568685	0.71\\
0.0001270081125	0.72\\
0.0002564174566	0.73\\
0.7071067812	0.93\\
0.7071067814	0.94\\
0.7071067818	0.95\\
0.7071067826	0.96\\
0.7071067882	0.97\\
1	0.99\\
1	1\\
};
\addlegendentry{Proposed}

\addplot[const plot, color=mycolor1, line width=1.5pt] table[row sep=crcr] {%
1.019303273e-05	0\\
1.019303273e-05	0.01\\
1.089649156e-05	0.02\\
1.111048715e-05	0.03\\
1.1807645e-05	0.04\\
1.303688368e-05	0.05\\
1.318161312e-05	0.06\\
1.352783612e-05	0.07\\
1.375088906e-05	0.08\\
1.442971461e-05	0.09\\
1.472987971e-05	0.1\\
1.515745196e-05	0.11\\
1.554613494e-05	0.12\\
1.614962978e-05	0.13\\
1.743435368e-05	0.14\\
1.752755095e-05	0.15\\
1.767524749e-05	0.16\\
1.811417966e-05	0.17\\
1.844726031e-05	0.18\\
1.951337553e-05	0.19\\
2.05216055e-05	0.2\\
2.116140005e-05	0.21\\
2.155766282e-05	0.22\\
2.255981219e-05	0.23\\
2.310616716e-05	0.24\\
2.423455967e-05	0.25\\
2.453643317e-05	0.26\\
2.50637845e-05	0.27\\
2.646852738e-05	0.28\\
2.726728388e-05	0.29\\
2.735923818e-05	0.3\\
3.230551794e-05	0.31\\
3.311117728e-05	0.32\\
3.538084089e-05	0.33\\
3.813183267e-05	0.34\\
4.376696726e-05	0.35\\
4.385404555e-05	0.36\\
4.563661833e-05	0.37\\
5.110439568e-05	0.38\\
5.195874839e-05	0.39\\
5.850909293e-05	0.4\\
5.866977821e-05	0.41\\
6.063911719e-05	0.42\\
6.390814832e-05	0.43\\
6.436427866e-05	0.44\\
7.083335948e-05	0.45\\
7.950073283e-05	0.46\\
8.764826914e-05	0.47\\
8.917497869e-05	0.48\\
9.034326226e-05	0.49\\
9.589803469e-05	0.5\\
9.856285731e-05	0.51\\
9.953657322e-05	0.52\\
0.0001091728164	0.53\\
0.0001433850015	0.54\\
0.0001474506889	0.55\\
0.7071067812	0.92\\
0.7071067814	0.93\\
0.7071067814	0.94\\
0.7071067817	0.95\\
0.7071067821	0.96\\
0.7071067827	0.97\\
0.7071067989	0.98\\
1	0.99\\
1.000000001	1\\
};
\addlegendentry{Baseline}

\end{axis}

\begin{axis}[%
width=1.166\columnwidth,
height=0.798\columnwidth,
at={(-0.152\columnwidth,-0.088\columnwidth)},
scale only axis,
xmin=0,
xmax=1,
ymin=0,
ymax=1,
axis line style={draw=none},
ticks=none,
axis x line*=bottom,
axis y line*=left
]
\end{axis}
\end{tikzpicture}%
        \caption{}
        \label{fig:GOSPAcdf_APdep}\vspace{-2pt}
    \end{subfigure}
    \vspace{-8pt}
  \caption{Comparison of the proposed framework with a fixed AP-mode benchmark. (a) Average image entropy versus average communication sum-SE for different values of $\beta$. (b) CDF of the GOSPA metric over $100$ different STO realizations for $\beta=0.75$.}
 \label{fig:Ent_GOSPA_APdep}
\end{figure}
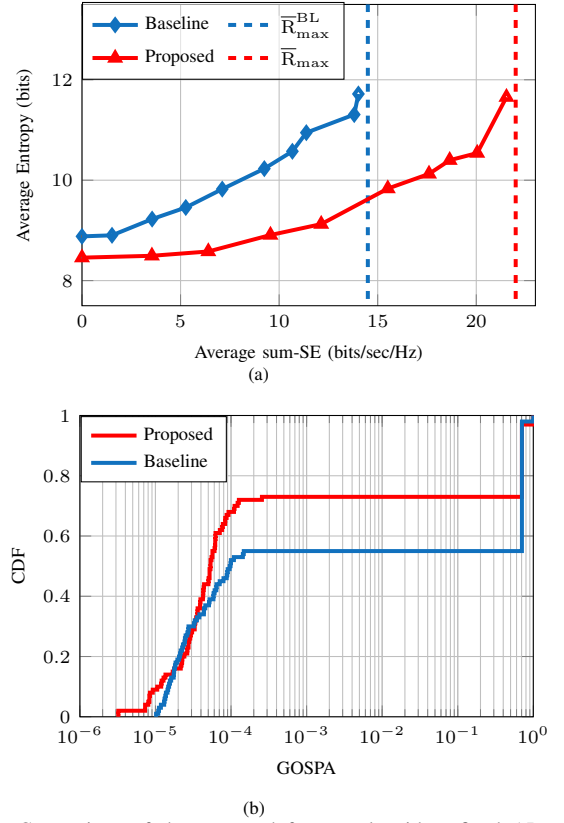

 \vspace{-8pt}
\subsection{Sensing-Communication Tradeoff}
We next characterize the sensing-communication trade-off achieved by the proposed framework. For each sum-SE budget $\beta$, the optimizer is executed over $B=100$ independent STO realizations. The image entropy and the sum-SE are averaged across all realizations. The trade-off curve is obtained by plotting the average image entropy against the average sum-SE for different $\beta$ values. We also plot the average of the maximum achievable sum-SE, $\overline{\mathrm{R}}_{\rm max}=1/B\sum_{b=1}^{B}\mathrm{R}_{\rm max}^{(b)},$
where $\mathrm{R}_{\rm max}^{(b)}=\mathrm{R}(\mathbf{b}=\mathbf{1},\mathbf{S}=0,\alpha=1)$ denotes the maximum sum-SE for the $b$-th STO realization. Finally, the CDF of the GOSPA metric is plotted over the same STO realizations.

\subsubsection{Impact of AP Mode Selection}\label{sec:AP_mode_sel_study}

We first compare the proposed framework with a baseline employing a fixed AP-mode configuration. In the baseline, $N=\lceil G/2\rceil$ APs are assigned to transmit mode and the remaining $M=G-N$ APs to receive mode. The APs in the grid are arranged in an alternating pattern, resulting in the AP-mode vector $\mathbf{b}_{\rm BL}=[0,1,0,\cdots]^\top$. It provides a sensing-oriented deployment with a large number of uniformly distributed bistatic pairs. Unlike the proposed framework, $\mathbf{b}_{\rm BL}$ is fixed during LTO and STO. During LTO, the sensing subcarriers are optimized using \eqref{eqn:interleaved} and \eqref{eqn:kappa_max}, and the power factor using \eqref{eqn:opt_alpha_LTO} for fixed AP mode $\mathbf{b}_{\rm BL}$. During STO, the sensing subcarrier allocation obtained from LTO, and the power factor are adapted using \eqref{eqn:opt_kappa_STO} and \eqref{eqn:alpha_opt} to satisfy the instantaneous sum-SE constraint. The average of the maximum sum-SEs across STO realizations for the baseline is computed as $\overline{\mathrm{R}}_{\rm max}^{\rm BL}=1/B\sum_{b=1}^{B}\mathrm{R}^{(b)}(\mathbf{b}_{\rm BL},\mathbf{S}=\mathbf{0},\alpha=1)$.

Fig.~\ref{fig:EntvsSE_APdep} plots the average image entropy against the average sum-SE for the proposed framework and the baseline. The proposed framework attains lower average image entropy while simultaneously achieving a higher average sum-SE. Fig.~\ref{fig:GOSPAcdf_APdep} shows the GOSPA CDF for $\beta=0.75$. The proposed framework achieves $\text{GOSPA}<10^{-4}$ in $72\%$ of the STO realizations, compared with $58\%$ for the baseline. These realizations correspond to correct detection and localization of all targets without missed detections or false alarms. The results demonstrate that a fixed AP-mode configuration  limits the achievable sensing-communication trade-off, whereas adaptive AP mode selection provides greater flexibility, improving imaging performance without sacrificing communication.

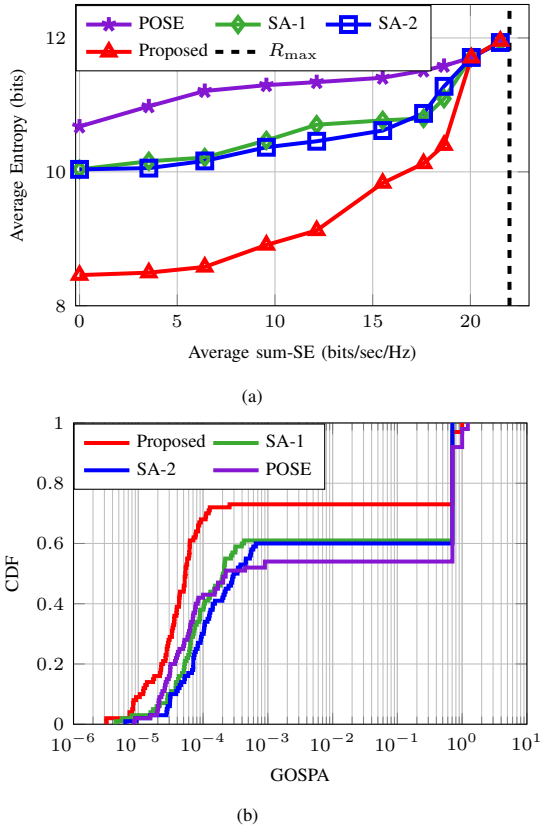
\begin{figure}[t]
    \centering
     \begin{subfigure}[h]{0.75\linewidth}
        \centering
%
\definecolor{mycolor1}{rgb}{0.23137,0.66667,0.19608}%
\definecolor{mycolor2}{rgb}{0.52157,0.08627,0.81961}%
\definecolor{mycolor3}{rgb}{0.12941,0.12941,0.12941}%
\begin{tikzpicture}

\begin{axis}[%
width=0.903\columnwidth,
height=0.6\columnwidth,
at={(0\columnwidth,0\columnwidth)},
scale only axis,
xmin=-0.2,
xmax=23,
xlabel style={font=\color{mycolor3},font=\scriptsize},
xlabel={Average sum-SE  (bits/sec/Hz)},
ymin=8,
ymax=12.5,
ylabel style={font=\color{mycolor3}, font=\scriptsize},
ylabel={Average Entropy (bits)},
axis background/.style={fill=white},
xmajorgrids,
ymajorgrids,
tick label style={font=\scriptsize},
legend style={    at={(0.00013,0.778)},  anchor=south west, 
    legend cell align=left,     align=left, 
    font=\scriptsize,     fill opacity=1, 
    draw opacity=1,     text opacity=1,
    legend columns=3 
}
]
\addplot [color=mycolor2, line width=1.5pt, mark size=3.8pt, mark=star,mark size=2.8pt, mark options={solid, mycolor2}]
  table[row sep=crcr]{%
0	10.67721\\
3.546512	10.97854121\\
6.408410469	11.207995123\\
9.56123	11.295321\\
12.13463564	11.3402632\\
15.51232	11.40521\\
17.60281881	11.512154\\
18.6521	11.5832\\
20.0482017	11.70721\\
21.5321	11.950721\\
};
\addlegendentry{POSE}

\addplot [color=mycolor1, line width=1.5pt, mark=diamond,mark size=2.8pt, mark options={solid, mycolor1}]
  table[row sep=crcr]{%
0	10.037830786\\
3.546512	10.158075306\\
6.408410469	10.21374984\\
9.56123	10.469605742\\
12.13463564	10.703960574\\
15.51232	10.76896057\\
17.60281881	10.79756057\\
18.6521	11.093452606\\
20.0482017	11.70893\\
21.5321	11.9523\\
};
\addlegendentry{SA-1}

\addplot [color=blue, line width=1.5pt, mark size=2.8pt, mark=square, mark options={solid, blue}]
  table[row sep=crcr]{%
0	10.03716\\
3.546512	10.05603\\
6.408410469	10.1632132\\
9.56123	10.367482\\
12.13463564	10.457032\\
15.51232	10.61666386\\
17.60281881	10.873166639\\
18.6521	11.277305\\
20.0482017	11.70893\\
21.5321	11.932\\
};
\addlegendentry{SA-2}

\addplot [color=red, line width=1.5pt, mark=triangle,mark size=2.8pt, mark options={solid, red}]
  table[row sep=crcr]{%
0	8.4575\\
3.546512	8.4944\\
6.408410469	8.5801009\\
9.56123	8.90789\\
12.13463564	9.12534729\\
15.51232	9.8328\\
17.60281881	10.12534729\\
18.6521	10.3981\\
20.0482017	11.704\\
21.5321	11.954\\
};
\addlegendentry{Proposed}

\addplot [color=black, dashed, line width=1.5pt]
  table[row sep=crcr]{%
22	8\\
22	12.5\\
};
\addlegendentry{$R_{\rm max}$}

\end{axis}

\end{tikzpicture}%
        \caption{}
         \label{fig:EntvsSEsub_aloc}
    \end{subfigure}
    \begin{subfigure}[h]{0.75\linewidth}
        \centering
%
\definecolor{mycolor1}{rgb}{0.23137,0.66667,0.19608}%
\definecolor{mycolor2}{rgb}{0.52100,0.08600,0.81900}%
\definecolor{mycolor3}{rgb}{0.12941,0.12941,0.12941}%
\begin{tikzpicture}

\begin{axis}[%
width=0.903\columnwidth,
height=0.6\columnwidth,
at={(0\columnwidth,0\columnwidth)},
scale only axis,
xmode=log,
xmin=1e-06,
xmax=10,
xminorticks=true,
xlabel style={font=\color{mycolor3}, font=\scriptsize},
xlabel={GOSPA},
ymin=0,
ymax=1,
ylabel style={font=\color{mycolor3}, font=\scriptsize},
ylabel={CDF},
axis background/.style={fill=white},
xmajorgrids,
xminorgrids,
ymajorgrids,
tick label style={font=\scriptsize},
legend style={    at={(0.00013,0.7782)},  anchor=south west, 
    legend cell align=left,     align=left, 
    font=\scriptsize,     fill opacity=1, 
    draw opacity=1,     text opacity=1,
    legend columns=2 
}
]
\addplot[const plot, color=red, line width=1.5pt] table[row sep=crcr] {%
3.22349505e-06	0\\
3.22349505e-06	0.01\\
3.227616839e-06	0.02\\
7.297711782e-06	0.04\\
8.042206057e-06	0.05\\
8.314285323e-06	0.07\\
8.450342008e-06	0.08\\
9.211811831e-06	0.09\\
1.05591633e-05	0.1\\
1.19125871e-05	0.11\\
1.206771134e-05	0.12\\
1.278483941e-05	0.13\\
1.356665652e-05	0.14\\
1.691814848e-05	0.15\\
1.762980643e-05	0.16\\
2.161630216e-05	0.17\\
2.21534807e-05	0.18\\
2.255713267e-05	0.19\\
2.265611892e-05	0.2\\
2.414281846e-05	0.21\\
2.635554909e-05	0.22\\
2.643025025e-05	0.23\\
2.741974537e-05	0.25\\
2.805875066e-05	0.26\\
2.888111242e-05	0.27\\
2.933321225e-05	0.28\\
3.056348516e-05	0.29\\
3.307371612e-05	0.31\\
3.372063262e-05	0.32\\
3.45682835e-05	0.33\\
3.583150535e-05	0.34\\
3.603543791e-05	0.35\\
3.610559673e-05	0.36\\
3.882215839e-05	0.37\\
3.907826637e-05	0.38\\
3.935180812e-05	0.39\\
4.250331792e-05	0.4\\
4.2813525e-05	0.41\\
4.285162651e-05	0.42\\
4.341724317e-05	0.43\\
4.396111856e-05	0.44\\
5.026355899e-05	0.45\\
5.040336441e-05	0.46\\
5.204700503e-05	0.49\\
5.350067844e-05	0.5\\
5.359422803e-05	0.52\\
5.514361973e-05	0.53\\
5.6784347e-05	0.55\\
6.052410115e-05	0.56\\
6.145877032e-05	0.57\\
6.180869603e-05	0.59\\
6.239452641e-05	0.6\\
6.241813353e-05	0.61\\
7.109910286e-05	0.62\\
7.632040234e-05	0.63\\
7.786973935e-05	0.64\\
8.36541538e-05	0.65\\
8.376212077e-05	0.66\\
8.738805592e-05	0.67\\
9.257583818e-05	0.68\\
0.0001094948643	0.69\\
0.0001114324411	0.7\\
0.0001222568685	0.71\\
0.0001270081125	0.72\\
0.0002564174566	0.73\\
0.7071067812	0.93\\
0.7071067814	0.94\\
0.7071067818	0.95\\
0.7071067826	0.96\\
0.7071067882	0.97\\
1	0.99\\
1	1\\
};
\addlegendentry{Proposed}

\addplot[const plot, color=mycolor1, line width=1.5pt] table[row sep=crcr] {%
4.451118757e-06	0\\
4.451118757e-06	0.01\\
5.549645593e-06	0.02\\
7.846794512e-06	0.03\\
1.543945469e-05	0.04\\
1.684853548e-05	0.05\\
1.915229024e-05	0.06\\
2.147442401e-05	0.07\\
2.805813023e-05	0.08\\
2.832161508e-05	0.09\\
2.862773374e-05	0.1\\
3.20407642e-05	0.11\\
3.765223519e-05	0.12\\
3.791638007e-05	0.13\\
3.805507876e-05	0.14\\
4.135691682e-05	0.15\\
4.216045779e-05	0.16\\
4.735703569e-05	0.17\\
5.02057823e-05	0.18\\
5.113782632e-05	0.19\\
5.173560531e-05	0.2\\
5.23266e-05	0.21\\
5.557182448e-05	0.22\\
5.889112488e-05	0.23\\
6.163714481e-05	0.24\\
6.179237417e-05	0.25\\
6.343304242e-05	0.26\\
6.367896351e-05	0.27\\
6.43830607e-05	0.28\\
6.478442137e-05	0.29\\
7.023118195e-05	0.3\\
7.171182691e-05	0.31\\
7.353353177e-05	0.32\\
7.512208862e-05	0.33\\
7.987913517e-05	0.34\\
8.780748795e-05	0.35\\
8.923821206e-05	0.36\\
8.978542821e-05	0.37\\
9.003961545e-05	0.38\\
0.0001014940894	0.39\\
0.0001032074284	0.4\\
0.0001068884672	0.41\\
0.0001214456407	0.42\\
0.0001279165805	0.43\\
0.0001292674665	0.44\\
0.0001473462867	0.45\\
0.000153546273	0.46\\
0.0001867056171	0.47\\
0.0001929639688	0.48\\
0.0001982103112	0.49\\
0.0002019450041	0.5\\
0.0002109455741	0.51\\
0.0002114304997	0.52\\
0.0002134466314	0.53\\
0.000216409329	0.54\\
0.0002192377862	0.55\\
0.0002694774535	0.56\\
0.0002836141646	0.57\\
0.0003159203948	0.58\\
0.0003160710462	0.59\\
0.0003966515252	0.6\\
0.0004295647763	0.61\\
0.7071067812	1\\
};
\addlegendentry{SA-1}

\addplot[const plot, color=blue, line width=1.5pt] table[row sep=crcr] {%
6.295987991e-06	0\\
6.295987991e-06	0.01\\
9.335282357e-06	0.02\\
1.52398665e-05	0.03\\
2.728422835e-05	0.04\\
2.746745273e-05	0.05\\
2.834289333e-05	0.06\\
3.071929562e-05	0.07\\
3.093094922e-05	0.08\\
3.106190356e-05	0.09\\
3.145066328e-05	0.1\\
3.851364246e-05	0.11\\
4.065360366e-05	0.12\\
4.33848023e-05	0.13\\
4.590721037e-05	0.14\\
5.139079005e-05	0.15\\
5.252263049e-05	0.16\\
5.905086645e-05	0.17\\
6.752682555e-05	0.18\\
7.035690661e-05	0.19\\
7.071067812e-05	0.22\\
7.219059382e-05	0.23\\
7.403451994e-05	0.24\\
7.733679524e-05	0.25\\
8.22590708e-05	0.26\\
8.319872495e-05	0.27\\
9.15846916e-05	0.28\\
9.305356352e-05	0.29\\
9.545145877e-05	0.3\\
0.0001023063236	0.31\\
0.0001036149461	0.32\\
0.0001050312783	0.33\\
0.0001067713463	0.34\\
0.0001172188277	0.35\\
0.0001242707107	0.36\\
0.0001263096046	0.37\\
0.0001299454518	0.38\\
0.000140773261	0.39\\
0.0001414126543	0.4\\
0.0001525697221	0.41\\
0.0001986201537	0.42\\
0.0002063852843	0.43\\
0.0002157071068	0.44\\
0.0002232828782	0.45\\
0.000247071068	0.46\\
0.0002695584597	0.47\\
0.0002812559232	0.48\\
0.0002846386655	0.49\\
0.0002958775348	0.5\\
0.0003405379828	0.51\\
0.0003444882088	0.52\\
0.0003782092996	0.53\\
0.0004552918911	0.54\\
0.0004570710678	0.55\\
0.000529111189	0.56\\
0.0005470710678	0.57\\
0.0005552011765	0.58\\
0.0005961010451	0.59\\
0.0006523707107	0.6\\
0.7071067812	0.96\\
0.7071067827	0.97\\
0.7071067851	0.98\\
0.7071074493	0.99\\
0.7071079838	1\\
};
\addlegendentry{SA-2}

\addplot[const plot, color=mycolor2, line width=1.5pt] table[row sep=crcr] {%
8.624827163e-06	0\\
8.624827163e-06	0.01\\
8.899678773e-06	0.02\\
1.533396038e-05	0.03\\
1.935733685e-05	0.04\\
1.989963478e-05	0.05\\
2.025448266e-05	0.06\\
2.059306307e-05	0.07\\
2.064304629e-05	0.08\\
2.12144682e-05	0.09\\
2.274892368e-05	0.1\\
2.35414941e-05	0.11\\
2.540983029e-05	0.12\\
2.588387472e-05	0.13\\
2.648532157e-05	0.14\\
2.757398563e-05	0.15\\
2.959274594e-05	0.16\\
3.064890039e-05	0.17\\
3.106645382e-05	0.18\\
3.116754902e-05	0.19\\
3.143853511e-05	0.2\\
3.68996652e-05	0.21\\
3.808364354e-05	0.22\\
3.958089204e-05	0.23\\
4.086931976e-05	0.24\\
4.329952379e-05	0.25\\
4.932625377e-05	0.26\\
5.099947919e-05	0.27\\
5.157631795e-05	0.28\\
5.735590351e-05	0.29\\
5.919864184e-05	0.3\\
6.069832202e-05	0.31\\
6.313066494e-05	0.32\\
6.344539389e-05	0.33\\
6.536723196e-05	0.34\\
6.933326301e-05	0.35\\
6.96400994e-05	0.36\\
7.116798372e-05	0.37\\
7.530811761e-05	0.38\\
7.619709836e-05	0.39\\
7.712710053e-05	0.4\\
8.255858854e-05	0.41\\
8.466146653e-05	0.42\\
0.0001000148356	0.43\\
0.0001511300227	0.44\\
0.00015406915	0.45\\
0.0001700845834	0.47\\
0.0002039182482	0.48\\
0.0002045831512	0.49\\
0.0002235588998	0.51\\
0.0004452481142	0.52\\
0.0009008622582	0.54\\
0.7071067812	0.87\\
0.7071067814	0.88\\
0.7071067839	0.89\\
0.7071067846	0.9\\
0.7071067873	0.91\\
0.7071067876	0.92\\
1	0.93\\
1	0.94\\
1	0.95\\
1.000000001	0.96\\
1.000000001	0.97\\
1.000000002	0.98\\
1.224744871	0.99\\
1.224744873	1\\
};
\addlegendentry{POSE}

\end{axis}

\end{tikzpicture}%
        \caption{}
        \label{fig:GOSPAcdfsub_aloc}
    \end{subfigure}    \vspace{-8pt}
  \caption{Comparison of sensing subcarrier allocation strategies: (a) Average image entropy versus the average communication sum-SE; (b) CDF of the GOSPA metric for different STO realizations for $\beta=0.75$. }
    \label{fig:sub_aloc}
\end{figure}

\subsubsection{Impact of Sensing Subcarrier Allocation}
We compare the performance of different sensing subcarrier allocation strategies in Fig.~\ref{fig:sub_aloc}. 
The proposed framework employs the maximally interleaved subcarrier allocation in \eqref{eqn:interleaved}. We compare it with two baseline strategies denoted by SA-1 and SA-2. In SA-1, each transmit AP is assigned a contiguous block of $\kappa$ subcarriers, and the blocks occupy consecutive frequency intervals. In SA-2, the block of $\kappa$ subcarriers assigned to each transmit AP are uniformly distributed across the available bandwidth. Mathematically, the sensing subcarriers assigned to the $n$-th transmit AP are
\begin{align}
    &\text{SA-1: } \mathcal{S}_{n,\rm sen} = \left\{(n-1)\kappa+1,\; (n-1)\kappa+2,\; \ldots,\; n\kappa\right\},\nonumber \\
    &\text{SA-2: } \mathcal{S}_{n,\rm sen} = \left\{ (n-1)(\kappa+\Delta) + 1,\; \ldots,\; n(\kappa+\Delta)-\Delta \right\},\nonumber
\end{align}
{where $\Delta=\left\lfloor{S-N\kappa}/{(N-1)}\right\rfloor$. We also consider the POSE configuration with $\kappa=1$.

We plot the average image entropy versus the average sum-SE by varying $\beta$ in Fig.~\ref{fig:EntvsSEsub_aloc}, and the GOSPA CDF with $\beta=0.75$ in Fig.~\ref{fig:GOSPAcdfsub_aloc}. Fig.~\ref{fig:EntvsSEsub_aloc} shows that the proposed interleaved allocation achieves the lowest average entropy over the sum-SE regime. Fig.~\ref{fig:GOSPAcdfsub_aloc} further shows that the proposed scheme correctly detects and localizes all targets, in $72\%$ of the STO realizations. The same is achieved in $60\%$ of the STO realizations for both SA-1 and SA-2, and in $57\%$ of the STO realizations for POSE. 
The proposed strategy performs best because each transmit AP spans a large effective bandwidth, improving their range resolution. In contrast, SA-1 and SA-2 use locally clustered subcarriers, limiting the bandwidth of each transmit AP. Although SA-2 distributes these clusters across the full bandwidth, its per-transmit AP bandwidth remains limited. This leads to performance similar to SA-1. POSE performs worst because only one sensing subcarrier per transmit AP is used, and therefore has no frequency diversity.

\vspace{-8pt}

\begin{figure}[t]
    \centering
     \begin{subfigure}[h]{0.75\linewidth}
        \centering
%
\definecolor{mycolor1}{rgb}{0.23137,0.66667,0.19608}%
\definecolor{mycolor2}{rgb}{0.52157,0.08627,0.81961}%
\definecolor{mycolor3}{rgb}{0.12941,0.12941,0.12941}%
\begin{tikzpicture}

\begin{axis}[%
width=0.903\columnwidth,
height=0.6\columnwidth,
at={(0\columnwidth,0\columnwidth)},
scale only axis,
xmin=-0.2,
xmax=23,
xlabel style={font=\color{mycolor3},font=\scriptsize},
xlabel={Average sum-SE  (bits/sec/Hz)},
ymin=8,
ymax=14,
ylabel style={font=\color{mycolor3}, font=\scriptsize},
ylabel={Average Entropy (bits)},
axis background/.style={fill=white},
xmajorgrids,
ymajorgrids,
tick label style={font=\scriptsize},
legend style={    at={(0.00013,0.77292)},  anchor=south west, 
    legend cell align=left,     align=left, 
    font=\scriptsize,     fill opacity=1, 
    draw opacity=1,     text opacity=1,
    legend columns=3 
}
]

\addplot [color=blue, line width=1.5pt, mark size=2.8pt, mark=square, mark options={solid, blue}]
  table[row sep=crcr]{%
0	9.9006\\
3.546512	10.2398\\
6.402121    10.3456\\
9.51643    10.8489\\
12.01643    11.4536\\
15.56521	12.0605\\
17.65413    12.321
18.6513      12.2373\\
20.04521     12.3565\\
21.8321	12.8565\\
};
\addlegendentry{Superposed}

\addplot [color=mycolor2, line width=1.5pt, mark size=3.8pt, mark=star,mark size=2.8pt, mark options={solid, mycolor2}]
  table[row sep=crcr]{%
0	10.67721\\
3.546512	10.97854121\\
6.408410469	11.207995123\\
9.56123	11.295321\\
12.13463564	11.3402632\\
15.51232	11.40521\\
17.60281881	11.512154\\
18.6521	11.5832\\
20.0482017	11.70721\\
21.5321	11.950721\\
};
\addlegendentry{POSE}

\addplot [color=red, line width=1.5pt, mark=triangle,mark size=2.8pt, mark options={solid, red}]
  table[row sep=crcr]{%
0	8.4575\\
3.546512	8.4944\\
6.408410469	8.5801009\\
9.56123	8.90789\\
12.13463564	9.12534729\\
15.51232	9.8328\\
17.60281881	10.12534729\\
18.6521	10.3981\\
20.0482017	11.704\\
21.5321	11.954\\
};
\addlegendentry{Proposed}

\addplot [color=black, dashed, line width=1.5pt]
  table[row sep=crcr]{%
22	8\\
22	14.5\\
};
\addlegendentry{$R_{\rm max}$}

\end{axis}

\end{tikzpicture}%
\vspace{-8pt}
        \caption{}
            \label{fig:Ent_SE_sup_orth}
    \end{subfigure}
        \begin{subfigure}[t]{0.75\linewidth}
        \centering
%
\definecolor{mycolor1}{rgb}{0.23137,0.66667,0.19608}%
\definecolor{mycolor2}{rgb}{0.52157,0.08627,0.81961}%
\definecolor{mycolor3}{rgb}{0.12941,0.12941,0.12941}%
\begin{tikzpicture}

\begin{axis}[%
width=0.903\columnwidth,
height=0.6\columnwidth,
at={(0\columnwidth,0\columnwidth)},
scale only axis,
xmode=log,
xmin=1e-06,
xmax=1.000000001,
xminorticks=true,
xlabel style={font=\color{mycolor2}, font=\scriptsize},
xlabel={GOSPA},
ymin=0,
ymax=1,
ylabel style={font=\color{mycolor2}, font=\scriptsize},
ylabel={CDF},
axis background/.style={fill=white},
xmajorgrids,
xminorgrids,
ymajorgrids,
tick label style={font=\scriptsize},
legend style={at={(0.00179,0.683)}, anchor=south west, legend cell align=left, align=left, font=\scriptsize}
]
\addplot[const plot, color=blue, line width=1.5pt] table[row sep=crcr] {%
9.407714742e-05	0\\
9.407714742e-05	0.01\\
9.864200526e-05	0.02\\
9.928219252e-05	0.03\\
0.0001144696969	0.04\\
0.0001224788173	0.05\\
0.0001273782954	0.06\\
0.0001310711127	0.07\\
0.0001332854099	0.08\\
0.0001334591561	0.09\\
0.0001359736843	0.1\\
0.0001410237837	0.11\\
0.0001430409775	0.12\\
0.0001592083756	0.13\\
0.0001625059283	0.14\\
0.000177467017	0.15\\
0.000185150253	0.17\\
0.0002026632451	0.19\\
0.0002075940447	0.2\\
0.0002084859413	0.21\\
0.0002149229811	0.23\\
0.000230400665	0.24\\
0.0002406337334	0.25\\
0.0002432626372	0.26\\
0.0002731334112	0.27\\
0.0002749275782	0.28\\
0.0002885138025	0.29\\
0.0002925035532	0.3\\
0.0003109852013	0.31\\
0.0003201701402	0.32\\
0.0003278195006	0.33\\
0.000330175446	0.34\\
0.0003499810223	0.35\\
0.0003570745333	0.36\\
0.0003600970042	0.37\\
0.0003647126611	0.38\\
0.0003659796777	0.39\\
0.0003696191181	0.4\\
0.0003951396888	0.41\\
0.0004115583108	0.42\\
0.0004516158931	0.43\\
0.0007071136643	0.44\\
0.0007453704357	0.45\\
0.0007607862205	0.46\\
0.0007651010075	0.47\\
0.001559161831	0.48\\
0.001643294245	0.49\\
0.001677607118	0.5\\
0.001919069475	0.51\\
0.002129582844	0.52\\
0.002232082518	0.53\\
0.003516610195	0.54\\
0.004035534766	0.55\\
0.004422259876	0.56\\
0.7071067812	0.83\\
0.7071074354	0.84\\
0.7071076557	0.85\\
0.7071079743	0.86\\
0.7071081544	0.87\\
0.707108524	0.88\\
0.707108569	0.89\\
0.7071093341	0.9\\
0.7071144461	0.91\\
0.7071164533	0.92\\
0.7071286752	0.93\\
0.7071504513	0.94\\
1.000002617	0.95\\
1.000004106	0.96\\
1.000013926	0.97\\
1.224745304	0.98\\
1.224750291	0.99\\
1.870842821	1\\
};
\addlegendentry{Superposed}

\addplot[const plot, color=red, line width=1.5pt] table[row sep=crcr] {%
4.149265216e-06	0\\
4.149265216e-06	0.02\\
8.043603044e-06	0.04\\
1.029672022e-05	0.05\\
1.032927287e-05	0.07\\
1.132051342e-05	0.08\\
1.148954412e-05	0.09\\
1.194567308e-05	0.1\\
1.227752428e-05	0.12\\
1.249863724e-05	0.13\\
1.303602341e-05	0.14\\
1.353647114e-05	0.15\\
1.373517563e-05	0.17\\
1.403165614e-05	0.18\\
1.488858095e-05	0.2\\
1.49530869e-05	0.22\\
1.561901023e-05	0.23\\
1.652830552e-05	0.24\\
1.671250234e-05	0.25\\
1.72904732e-05	0.27\\
1.800316663e-05	0.29\\
1.816431677e-05	0.3\\
1.961037225e-05	0.31\\
1.965612372e-05	0.32\\
1.970396603e-05	0.33\\
1.990901248e-05	0.34\\
2.196755413e-05	0.35\\
2.211571914e-05	0.36\\
2.261005416e-05	0.37\\
2.420151321e-05	0.39\\
2.698917715e-05	0.4\\
2.839912622e-05	0.41\\
2.854231959e-05	0.42\\
3.157917089e-05	0.43\\
3.177271936e-05	0.44\\
3.205018109e-05	0.45\\
3.205778608e-05	0.46\\
3.297019559e-05	0.47\\
3.521632085e-05	0.48\\
3.542200339e-05	0.49\\
3.839952793e-05	0.5\\
4.322063063e-05	0.51\\
4.423632302e-05	0.53\\
4.461809459e-05	0.55\\
4.47910899e-05	0.57\\
4.709657221e-05	0.58\\
4.713312408e-05	0.59\\
4.932837569e-05	0.6\\
6.072575893e-05	0.61\\
6.147976351e-05	0.63\\
7.567303583e-05	0.64\\
7.694201017e-05	0.65\\
8.333006159e-05	0.66\\
8.695977545e-05	0.67\\
0.0001800974893	0.69\\
0.7071067812	0.95\\
0.7071067812	0.96\\
0.7071067813	0.97\\
0.7071067814	0.99\\
0.7071067816	1\\
};
\addlegendentry{Proposed}

\addplot[const plot, color=mycolor2, line width=1.5pt] table[row sep=crcr] {%
8.624827163e-06	0\\
8.624827163e-06	0.01\\
8.899678773e-06	0.02\\
1.533396038e-05	0.03\\
1.935733685e-05	0.04\\
1.989963478e-05	0.05\\
2.025448266e-05	0.06\\
2.059306307e-05	0.07\\
2.064304629e-05	0.08\\
2.12144682e-05	0.09\\
2.274892368e-05	0.1\\
2.35414941e-05	0.11\\
2.540983029e-05	0.12\\
2.588387472e-05	0.13\\
2.648532157e-05	0.14\\
2.757398563e-05	0.15\\
2.959274594e-05	0.16\\
3.064890039e-05	0.17\\
3.106645382e-05	0.18\\
3.116754902e-05	0.19\\
3.143853511e-05	0.2\\
3.68996652e-05	0.21\\
3.808364354e-05	0.22\\
3.958089204e-05	0.23\\
4.086931976e-05	0.24\\
4.329952379e-05	0.25\\
4.932625377e-05	0.26\\
5.099947919e-05	0.27\\
5.157631795e-05	0.28\\
5.735590351e-05	0.29\\
5.919864184e-05	0.3\\
6.069832202e-05	0.31\\
6.313066494e-05	0.32\\
6.344539389e-05	0.33\\
6.536723196e-05	0.34\\
6.933326301e-05	0.35\\
6.96400994e-05	0.36\\
7.116798372e-05	0.37\\
7.530811761e-05	0.38\\
7.619709836e-05	0.39\\
7.712710053e-05	0.4\\
8.255858854e-05	0.41\\
8.466146653e-05	0.42\\
0.0001000148356	0.43\\
0.0001511300227	0.44\\
0.00015406915	0.45\\
0.0001700845834	0.47\\
0.0002039182482	0.48\\
0.0002045831512	0.49\\
0.0002235588998	0.51\\
0.0004452481142	0.52\\
0.0009008622582	0.54\\
0.7071067812	0.87\\
0.7071067814	0.88\\
0.7071067839	0.89\\
0.7071067846	0.9\\
0.7071067873	0.91\\
0.7071067876	0.92\\
1	0.93\\
1	0.94\\
1	0.95\\
1.000000001	0.96\\
1.000000001	0.97\\
1.000000002	0.98\\
1.224744871	0.99\\
1.224744873	1\\
};
\addlegendentry{POSE}

\end{axis}

\end{tikzpicture}%
\vspace{-8pt}
        \caption{}
        \label{fig:GOSPA_sup_orth}
    \end{subfigure}\vspace{-10pt}
  \caption{Comparison between the proposed orthogonal, superposition, and POSE schemes: (a) sensing-communication trade-off by plotting average image entropy versus average sum-SE for different values of $\beta$; (b) CDF of GOSPA values over $100$ STO realizations for $\beta=0.75$.}
\end{figure}
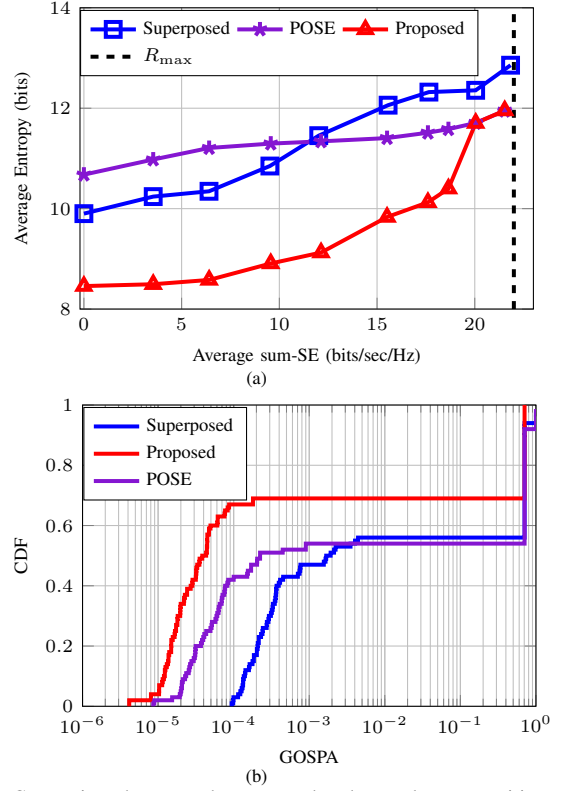
\subsection{Comparison with Superposition Scheme}\vspace{-3pt}
We compare the proposed framework with the superposition based ISAC scheme in~\cite{tagliaferri2025integratingphasecoherentmultistaticimaging}. Unlike the proposed framework, which assigns orthogonal sensing and communication subcarriers, the superposition scheme transmits both signals over all $S$ subcarriers. Each transmit AP superposes pseudo-orthogonal sensing signals with the communication signals using power-splitting factor $\alpha$. This results in mutual sensing-communication interference. The proposed scheme eliminates this~interference~through~orthogonal~subcarrier~allocation.

{Fig.~\ref{fig:Ent_SE_sup_orth} plots the average image entropy versus the average sum-SE over $100$ STO realizations for different $\beta$. The proposed framework consistently achieves a better sensing-communication trade-off than the superposition and POSE schemes. Furthermore, the superposition scheme outperforms POSE in the low-sum-SE regime, whereas POSE achieves a better trade-off at higher sum-SEs. At low sum-SE requirements $(\beta<0.5)$, the superposition scheme benefits from transmitting sensing signals over all $S$ subcarriers. At higher sum-SE requirements $(\beta>0.5)$, higher fraction of transmit power is allocated to communication, reducing the sensing power. In superposition scheme, the stronger communication signal interferes with sensing, as both functions share the same subcarriers, further degrading the image quality. In contrast, POSE uses orthogonal sensing and communication subcarriers, avoiding this interference. Since POSE already operates with the minimum sensing-subcarrier allocation, $\kappa=1$, its imaging performance degrades more mildly in the high-SE regime.} Fig.~\ref{fig:GOSPA_sup_orth} shows the GOSPA CDFs for $\beta=0.75$. The proposed scheme correctly detects and localizes all targets in $72\%$ of the STO realizations, compared with $58\%$ for both the superposition scheme and POSE. Even when all targets are detected, the superposition scheme has larger localization errors than the orthogonal scheme. This is because sensing-communication interference distorts the reconstructed target peaks, reducing localization accuracy despite successful detection. 

\begin{figure}[t]
    \centering
    \begin{subfigure}[t]{0.75\linewidth}
        \centering
        \definecolor{mycolor1}{rgb}{0.12941,0.12941,0.12941}

\begin{tikzpicture}

\begin{axis}[
    width=0.903\columnwidth,
    height=0.6\columnwidth,
    scale only axis,
    xmin=0, xmax=6500,
    xlabel={Runtime (sec)},
    xlabel style={font=\color{mycolor1}\scriptsize},
    ylabel={Entropy},
    ylabel style={font=\color{blue}\scriptsize},
    axis y line*=left, 
    every y tick label/.append style={font=\color{blue}\scriptsize},
    every y tick/.append style={blue},
    axis background/.style={fill=white},
    xmajorgrids,
    ymajorgrids,
    tick label style={font=\scriptsize},
    legend style={at={(0.44809251399, 0.6006568)}, anchor=south west, legend cell align=left, align=left, font=\scriptsize},
]

\addplot [color=blue, line width=1.5pt, mark size=2.3pt, mark=triangle, mark options={solid, blue}]
  table[row sep=crcr]{%
    0 12.586\\323.3767 12.15823\\ 659.714 12.05922\\ 955.47 11.85850\\ 1.2817e+03 11.65887\\
    1.6029e+03 11.4522\\ 1.9078e+03 11.35846\\ 2.2215e+03 11.15886\\ 2.5429e+03 11.105824\\  2.8967e+03 11.0095102\\
    3.2424e+03 10.95593\\ 3.5469e+03 10.875599\\ 3.8952e+03 10.75642\\ 4.2205e+03 10.60915598\\ 4.5379e+03 10.5708\\
   4.8609e+03 10.45651\\ 5.1789e+03 10.31\\ 5.4957e+03 10.28\\ 5.8286e+03 10.28\\ 6.1607e+03 10.26\\
   6.4607e+03 10.25\\
};
\addlegendentry{Entropy ($\beta=0.75$)}

\addplot [color=blue, line width=1.5pt, mark size=2.3pt, mark=square, mark options={solid, blue}]
  table[row sep=crcr]{%
    0 12.590\\3.317029854009523e+02 11.40355\\ 6.325878212124153e+02 11.1470\\ 9.495653967515603e+02 10.810593\\ 1.265150926335487e+03 10.690714\\   1.586342993727142e+03 10.490766\\ 1.907471579693516e+03 10.350782\\2.216822780520271e+03 10.210678\\2.534522267558309e+03 10.10643\\ 2.850873477366218e+03 9.950521\\    3.175150550241505e+03 9.90373\\ 3.504083206931900e+03 9.850433\\ 3.833175939908044e+03 9.789\\ 4.142539411688157e+03 9.65\\ 4.461313002599461e+03 9.56\\
   4.767171832163307e+03 9.4\\ 5.074036824748439e+03 9.36\\ 5.391968331467406e+03 9.35\\ 5.725294634550253e+03 9.33\\ 6.035597975412717e+03 9.33\\
   6.357311763540317e+03 9.32\\
};
\addlegendentry{Entropy ($\beta=0.5$)}

\addlegendimage{color=red,  line width=1.5pt, mark=triangle, mark size=2.3pt}
\addlegendentry{Sum-SE  ($\beta=0.75$)}
\addlegendimage{color=red,line width=1.5pt, mark=square, mark size=2.3pt}
\addlegendentry{Sum-SE  ($\beta=0.5$)}
\end{axis}

\begin{axis}[
    width=0.903\columnwidth,
    height=0.6\columnwidth,
    scale only axis,
    xmin=0, xmax=6500,
    ymin=9.6, ymax=22,
    ylabel={Sum-SE  (bits/sec/Hz)},
    ylabel style={font=\color{red}\scriptsize},
    axis y line*=right, 
    axis x line=none,    
    every y tick label/.append style={font=\color{red}\scriptsize},
    every y tick/.append style={red},
]

\addplot [color=red, line width=1.5pt, mark=triangle, mark size=2.3pt, mark options={solid, red}]
  table[row sep=crcr]{%
    0 21.586\\323.3767 16.5823\\ 659.714 16.5922\\ 955.47 16.5850\\ 1.2817e+03 16.5887\\
    1.6029e+03 16.522\\ 1.9078e+03 16.5846\\ 2.2215e+03 16.5886\\ 2.5429e+03 16.5824\\  2.8967e+03 16.5102\\
    3.2424e+03 16.5593\\ 3.5469e+03 16.5599\\ 3.8952e+03 16.5642\\ 4.2205e+03 16.5598\\ 4.5379e+03 16.5708\\
   4.8609e+03 16.5651\\ 5.1789e+03 16.5600\\ 5.4957e+03 16.50596\\ 5.8286e+03 16.5658\\ 6.1607e+03 16.5887\\
   6.4607e+03 16.5562\\
};

\addplot [color=red, line width=1.5pt, mark size=2.3pt, mark=square, mark options={solid, red}]
  table[row sep=crcr]{%
    0 21.586\\3.317029854009523e+02 12.0355\\ 6.325878212124153e+02 12.1470\\ 9.495653967515603e+02 12.0593\\ 1.265150926335487e+03 12.0714\\   1.586342993727142e+03 12.0766\\ 1.907471579693516e+03 12.0782\\2.216822780520271e+03 12.0678\\2.534522267558309e+03 12.0643\\ 2.850873477366218e+03 12.0521\\    3.175150550241505e+03 12.0373\\ 3.504083206931900e+03 12.0433\\ 3.833175939908044e+03 12.0389\\ 4.142539411688157e+03 12.0225\\ 4.461313002599461e+03 12.0362\\
   4.767171832163307e+03 12.0225\\ 5.074036824748439e+03 12.0433\\ 5.391968331467406e+03 12.0521\\ 5.725294634550253e+03 12.0714\\ 6.035597975412717e+03 12.0593\\
   6.357311763540317e+03 12.0225\\
};

\end{axis}
\end{tikzpicture}\vspace{-8pt}
        \caption{}
        \label{fig:conv_LTO}
    \end{subfigure}
    \begin{subfigure}[t]{0.75\linewidth}
        \centering
        \input{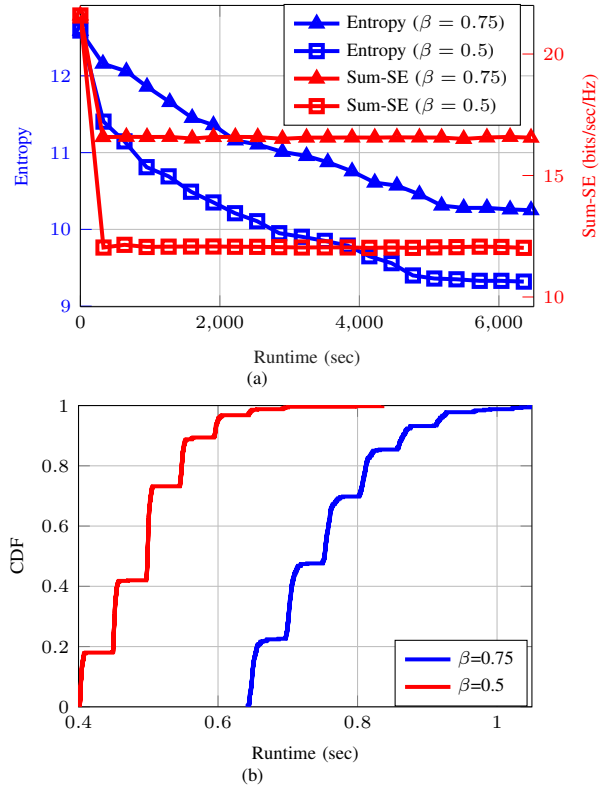}\vspace{-8pt}
        \caption{}
        \label{fig:conv_STO}
    \end{subfigure}
    \vspace{-10pt}
   \caption{ (a) Convergence of the LTO algorithm: Image entropy and sum-SE versus cumulative runtime for $\beta=\{0.75,0.5\}$. (b) CDF of the runtime of the STO algorithm for $\beta=\{0.75,0.5\}$.}
 \label{fig:Ent_vs_N}
\end{figure}
\begin{figure}[t]
    \centering
         \begin{subfigure}[t]{0.75\linewidth}
        \centering
         \includegraphics[width=\linewidth]{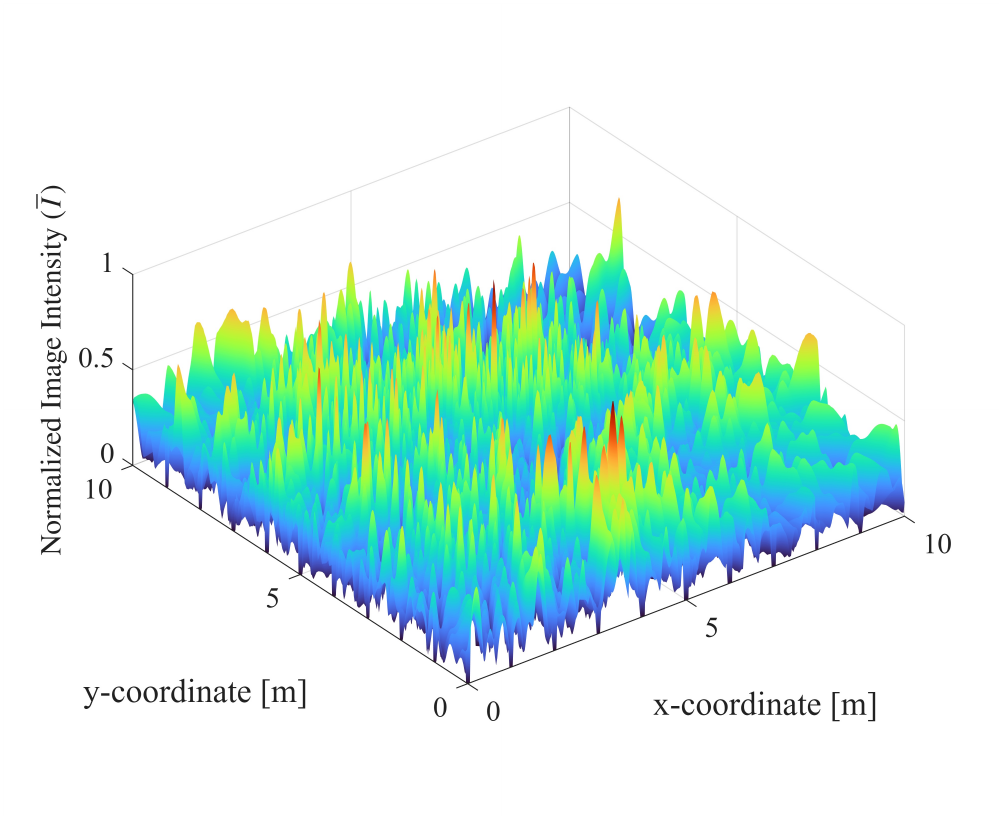}
        \vspace{-28pt}
         \caption{}
         \label{fig:Image_C}
    \end{subfigure}\vspace{0pt}
     \begin{subfigure}[t]{0.75\linewidth}
        \centering
         \includegraphics[width=\linewidth]{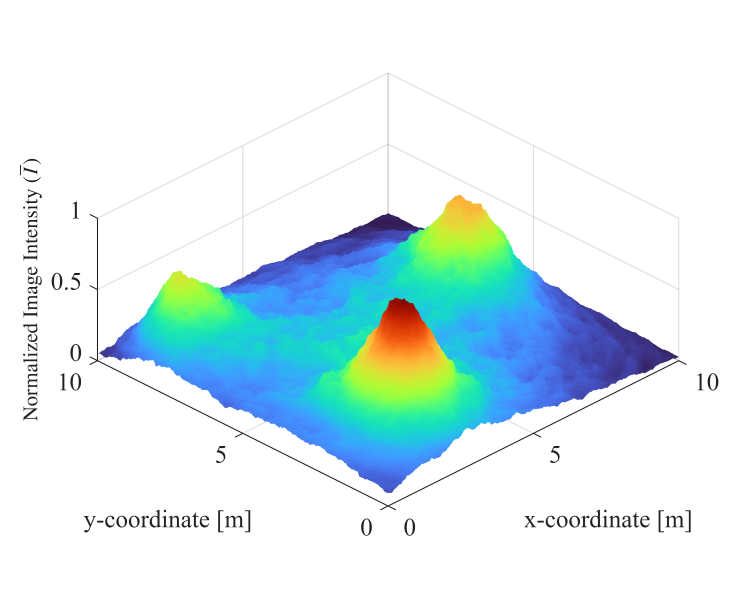}
         \vspace{-28pt}
         \caption{}
             \label{fig:Image_NC}
    \end{subfigure}\vspace{-8pt}
    \caption{Normalized image intensity under phase synchronization errors ($\phi_{\rm max}=\pi/2$ rad) for a representative STO realization. (a) Coherent imaging. (b) Non-coherent imaging. }
    \label{fig:Im_noisy}
\end{figure}
\vspace{-10pt}
{\subsection{Convergence and Complexity Analysis}
We next investigate the convergence and complexity of the LTO (Algorithm~\ref{algorithm:long_timescale}).  Fig.~\ref{fig:conv_LTO} plots the entropy and sum-SE of the synthetic scenario versus the runtime for $\beta=\{0.75,0.5\}$. For both values of $\beta$, the entropy decreases monotonically and converges within approximately $15$ iterations, corresponding to a runtime of about $5\times10^3$~s. 
The sum-SE is initially close to $\mathrm{R}_{\max}$ since the LTO is initialized at a communication-favorable operating point. It then drops in the first iteration to satisfy the prescribed sum-SE budget and remains constant.}
 
{Fig.~\ref{fig:conv_STO} shows the CDF of the STO runtime in Algorithm~\ref{algorithm:short_timescale} for $\beta=\{0.5,0.75\}$. The runtime CDF exhibits a staircase behavior because the sum-SE feasibility-restoration stage performs a discrete number of resource-allocation updates. Depending on the channel and UE geometry of each STO realization, the algorithm may reduce $\kappa$ several times and then switch one or more receive APs to transmit mode. Realizations requiring the same number of updates have similar runtimes, producing visible jumps in the CDF. A larger $\beta$ imposes a stricter sum-SE constraint, making the initial LTO solution more likely to be infeasible and requiring more restoration iterations. It also results in a larger communication subcarrier set, increasing the cost of each sum-SE evaluation. Hence, the runtime distribution shifts toward higher values as $\beta$ increases.}

\vspace{-8pt}
\subsection{Performance Evaluation for Phase Unsynchronized APs}
{We next evaluate the sensing and communication performance of the framework under phase synchronization errors across the APs. During LTO, the AP modes and sensing subcarrier allocation are optimized assuming perfectly synchronized APs in the synthetic scenario. During STO, the APs are phase-unsynchronized. Specifically, each transmit/receive AP is assumed to introduce an independent phase error $\{\phi_g\}_{g=1}^{G}$, where $\phi_g\sim\mathcal{U}(-\phi_{\rm max},\phi_{\rm max})$, to its transmit/receive signal.}

Fig.~\ref{fig:Im_noisy} shows the normalized image for the representative STO realization as in Fig.~\ref{fig:Im_opt}, assuming phase-unsynchronized APs with $\phi_{\rm max}=\pi/2$ rad. Fig.~\ref{fig:Image_C} corresponds to coherent imaging (Section~\ref{sec:coherent_imaging}), whereas Fig.~\ref{fig:Image_NC} corresponds to non-coherent imaging (Section~\ref{sec:noncoherent_imaging}). The coherent image in Fig.~\ref{fig:Image_C} with entropy $12.09$ bits, is severely degraded, as compared to Fig.~\ref{fig:Im_opt}, by numerous spurious peaks. This is because coherent imaging relies on inter-AP phase coherence, which is disrupted here by synchronization errors. In contrast, the non-coherent image in Fig.~\ref{fig:Image_NC}, which does not rely on inter-AP phase coherence, produces clean peaks at the target locations with entropy $11.21$ bits. However, compared with the coherent image with perfectly synchronized APs in Fig.~\ref{fig:Im_opt}, the peaks have reduced sharpness.

\begin{figure}[t]
    \centering
    \hspace{-50pt}
     \begin{subfigure}[h]{0.9\linewidth}
        \centering
        \definecolor{mycolor1}{rgb}{0.12941,0.12941,0.12941}

\begin{tikzpicture}

\begin{axis}[
    width=0.903\columnwidth,
    height=0.5\columnwidth,
    scale only axis,
    xmin=0, xmax=3.14,
    xlabel={Maximum Phase Error $\phi_{\rm max}$ (rad)},
    xlabel style={font=\color{mycolor1}\scriptsize},
    ymin=8, ymax=13,
    ylabel={Entropy},
    ylabel style={font=\scriptsize},
    axis y line*=left, 
    every y tick label/.append style={font=\scriptsize},
    every y tick/.append style={blue},
    axis background/.style={fill=white},
    xmajorgrids,
    ymajorgrids,
    tick label style={font=\scriptsize},
legend style={
    at={(0.0,0.27)},
    anchor=north west,
    legend columns=2,
    legend cell align=left,
    align=left,
    font=\tiny,
    fill=white,
    fill opacity=0.95,
    text opacity=1
},
]

\addplot [color=blue, dashdotted, line width=1.5pt]
  table[row sep=crcr]{%
0.001	11.628\\
0.005	11.628\\
0.01	11.628\\
0.05	11.628\\
0.1	    11.628\\
0.25    11.628\\
0.5	    11.628\\
1	    11.628\\
1.57	11.628\\
3.14	11.628\\
};
\addlegendentry{Entropy (Non-coherent,  $\beta=0.75$)}

\addplot [color=red, dashdotted, line width=1.5pt]
  table[row sep=crcr]{%
0.0002	        11.121\\
0.006135923152	11.121\\
0.0122718463	11.121\\
0.02454369261	11.121\\
0.04908738521	11.121\\
0.09817477042	11.121\\
0.1963495408	11.121\\
0.3926990817	11.121\\
0.7853981634	11.121\\
1.570796327	    11.121\\
3.141592654	    11.121\\
};
\addlegendentry{Entropy (Non-coherent,  $\beta=0.5$)}

\addplot [color=red, line width=1.5pt, mark=square]
  table[row sep=crcr]{%
0.001	9.7382\\
0.1	    9.81504\\
0.25	10.00969\\
0.5	    10.29\\
0.8     10.84\\
1	    11.048\\
1.57	11.53\\
2.09    11.9\\
2.6180  12.05\\   
3.14	12.29\\
};
\addlegendentry{Entropy (Coherent,  $\beta=0.5$)}

\addplot [color=blue, line width=2pt, mark=diamond]
  table[row sep=crcr]{%
0.001	10.17382\\
0.1	    10.37\\
0.25	10.8\\
0.5	    11.03453\\
0.8     11.32\\
1	    11.88\\
1.57	12.03\\
2.09    12.34\\
2.6180  12.36\\   
3.14	12.6\\
};
\addlegendentry{Entropy (Coherent,  $\beta=0.75$)}

\addlegendimage{color=black,  line width=1.5pt, mark=diamond, mark size=2pt}
\addlegendentry{Sum-SE  ($\beta=0.75$)}
\addlegendimage{color=black,line width=1.5pt, mark=square, mark size=1.5pt}
\addlegendentry{Sum-SE  ($\beta=0.5$)}
\end{axis}

\begin{axis}[
    width=0.903\columnwidth,
    height=0.5\columnwidth,
    scale only axis,
    xmin=0, xmax=3.14,
    ymin=5, ymax=17,
    ylabel={Average sum-SE (bits/sec/Hz)},
    ylabel style={font=\color{black}\scriptsize},
    axis y line*=right, 
    axis x line=none,    
    every y tick label/.append style={font=\scriptsize},
    every y tick/.append style={black},
]

\addplot [color=black, line width=1.5pt, mark=square]
  table[row sep=crcr]{%
0.001	11\\
0.1	    10.95\\
0.25	10.9\\
0.5	    10.92\\
0.8     10.84\\
1	    10.81\\
1.57	10.42\\
2.09    9.02\\
2.6180  8.39\\   
3.14	7.93\\
};

\addplot [color=black, line width=2pt, mark=diamond]
  table[row sep=crcr]{%
0.001	16.5\\
0.1	    16.504\\
0.25	16.5\\
0.5	    16.43\\
0.8     16.31\\
1	    16.24\\
1.57	15.6\\
2.09    13.513\\
2.6180  12.59\\   
3.14	11.895\\
};

\end{axis}
\end{tikzpicture}
    \end{subfigure}\vspace{-8pt}
  \caption{Average image entropy (coherent and non-coherent imaging), and average sum-SE across different STO realizations versus maximum phase~error~$\phi_{\rm max}$.}
    \label{fig:phase_noise}
\end{figure}
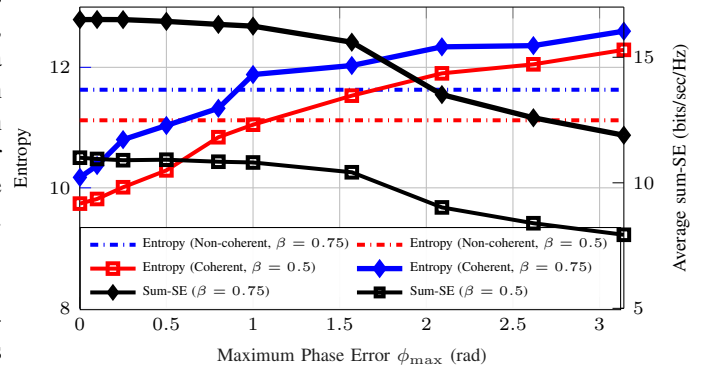

{We next plot the average image entropy for coherent and non-coherent imaging, and the average sum-SE over the STO realizations versus the maximum phase error $\phi_{\rm max}$  in Fig.~\ref{fig:phase_noise}. The system is optimized for $\beta=\{0.5,0.75\}$. For both $\beta$ values, the entropy of coherent imaging increases with $\phi_{\rm max}$, whereas the entropy of non-coherent imaging remains constant. This is because non-coherent imaging does not rely on phase coherence across APs. Coherent imaging achieves lower entropy for small phase errors, but its performance degrades beyond non-coherent imaging for $\phi_{\rm max}>1$ rad, indicating the regime in which coherent processing remains beneficial. For both $\beta$ values, the average sum-SE decreases as $\phi_{\rm max}$ increases. This is because the phase errors distort the coherent combining of the distributed AP signals at the UEs, reducing the intended beamforming gain.}

\vspace{-2pt}
\section{Conclusion}
\vspace{-2pt}
In this paper, we proposed an orthogonal subcarrier–based ISAC framework for phase-coherent D-MIMO systems, enabling joint communication and multi-static sensing through radio imaging. By separating sensing and communication in the frequency domain and distributing sensing subcarriers across APs, the proposed design eliminates inter-function interference.  
We develop a two-timescale optimization framework that jointly selects AP modes, sensing subcarriers, and sensing–communication power splitting, minimizing the   entropy of the reconstructed image under communication constraints. 
Simulation results demonstrate that the proposed orthogonal design consistently outperforms conventional superposed coding in terms of sensing–communication trade-offs, achieving lower entropy and improved localization accuracy across a wide range of operating conditions. We also investigate the robustness of coherent imaging with unsynchronized APs, identifying the transition point beyond which non-coherent processing becomes preferable. Overall, the proposed framework offers a high-resolution ISAC solution for distributed MIMO systems, advancing toward reliable environmental awareness in future 6G networks.

\bibliographystyle{IEEEtran}
\bibliography{references, references_imported}
\end{document}